\documentclass[final,5p,times,twocolumn,authoryear]{elsarticle}

\usepackage{amssymb}
\usepackage{amsmath}

\usepackage[colorlinks,allcolors=blue]{hyperref}
\usepackage{rotating}
\usepackage{pdflscape}

\journal{Planetary and Space Science}

\begin{document}

\begin{frontmatter}



\title{Deep machine learning for meteor monitoring: advances with transfer learning and gradient-weighted class activation mapping}


\author[inst1,inst2]{Eloy Peña Asensio}
\author[inst2,inst3]{Josep M. Trigo-Rodríguez}
\author[inst2]{Pau Grèbol-Tomàs}
\author[inst4]{David Regordosa-Avellana}
\author[inst1]{Albert Rimola}

\affiliation[inst1]{organization={Departament de Química, Universitat Autònoma de Barcelona},
            addressline={Carrer dels Til·lers}, 
            city={Bellaterra},
            postcode={08193}, 
            state={Catalonia},
            country={Spain}}

\affiliation[inst2]{organization={Institut de Ciències de l’Espai (ICE, CSIC)},
            addressline={Campus UAB, C/ de Can Magrans S/N}, 
            city={Cerdanyola del Vallès},
            postcode={08193}, 
            state={Catalonia},
            country={Spain}}
            
\affiliation[inst3]{organization={Institut  d’Estudis  Espacials  de  Catalunya  (IEEC)},
            addressline={Carrer del Gran Capità, 2}, 
            city={Barcelona},
            postcode={08034}, 
            state={Catalonia},
            country={Spain}}

\affiliation[inst4]{organization={Spanish Meteor Network (SPMN, CSIC) },
            addressline={Campus UAB, C/ de Can Magrans S/N}, 
            city={Cerdanyola del Vallès},
            postcode={08193}, 
            state={Catalonia},
            country={Spain}}

\begin{abstract}

In recent decades, the use of optical detection systems for meteor studies has increased dramatically, resulting in huge amounts of data being analyzed. Automated meteor detection tools are essential for studying the continuous meteoroid incoming flux, recovering fresh meteorites, and achieving a better understanding of our Solar System. Concerning meteor detection, distinguishing false positives between meteor and non-meteor images has traditionally been performed by hand, which is significantly time-consuming. To address this issue, we developed a fully automated pipeline that uses Convolutional Neural Networks (CNNs) to classify candidate meteor detections. Our new method is able to detect meteors even in images that contain static elements such as clouds, the Moon, and buildings. To accurately locate the meteor within each frame, we employ the Gradient-weighted Class Activation Mapping (Grad-CAM) technique. This method facilitates the identification of the region of interest by multiplying the activations from the last convolutional layer with the average of the gradients across the feature map of that layer. By combining these findings with the activation map derived from the first convolutional layer, we effectively pinpoint the most probable pixel location of the meteor. We trained and evaluated our model on a large dataset collected by the Spanish Meteor Network (SPMN) and achieved a precision of 98\%. Our new methodology presented here has the potential to reduce the workload of meteor scientists and station operators and improve the accuracy of meteor tracking and classification.

\end{abstract}




\end{frontmatter}


\section{Introduction}

Meteors, popularly known as shooting stars, particularly the most luminous ones called fireballs or bolides, are spectacular physical processes that have fascinated mankind for centuries \citep{Trigo2022}. These dazzling streaks of light occur when a meteoroid enters the Earth's atmosphere at hypersonic velocity, causing intense heating through repeated collisions with air molecules \citep{Ceplecha1998SSRv, Silber2018AdSpR, Trigo2019hmep}. Meteors are formed due to the extreme heat produced by the interaction with the gaseous environment and the rising atmospheric pressure which causes the meteoroid to undergo rapid vaporization, a process known as ablation. This ablation leads to the formation of a luminous trail composed of ionized gas and fragmented debris, which can be observed and recorded from the ground using optical devices. Current digital video imagery provides a sequential recording useful to obtain complete light curves, high temporal and spatial resolution measurements, and spectra \citep{Hughes1978book, Koschny2017PSS, Subasinghe2017PSS, Drolshagen2021AA}. 

Traditionally, two classes of meteors are considered: those that are expected to occur in a specific period of the year because they are associated with meteoroid streams, and those that are sporadic and have no discernible periodic pattern \citep{Wiegert2004EMP, Jopek2013MNRAS, Dumitru2017AA, Jenniskens2017PSS, Vaubaillon2019, Eloy2022AJ, Eloy2023MNRAS}. Although showers exhibit regular activity, the sporadic meteors require constant sky monitoring to quantify the meteoroid flux and properties of the different sources \citep{Trigo2022MNRAS}.

Meteors can provide valuable information about the composition, dynamics, and origin of our Solar System \citep{Koschny2019SSRv}. By analyzing the physical and chemical properties of meteorites, which are pieces of meteoroids that survive their journey through the Earth's atmosphere and land on the surface, scientists can gain insight into the formation and evolution of comets and asteroids, as they provide information about the age of the Solar System, the composition of the early solar nebula, and the processes that led to the formation of the planets \citep{Bottke2002, Walker2006book}.

The growing interest in meteoritics has driven an increase in the number of video meteor detection networks around the world \citep{Ceplecha1987, Ceplecha1998SSRv, Koten2019msme, Colas2020AA}. Comprised of strategically placed stations equipped with cameras and other sensors, these networks are designed to monitor atmospheric volumes with clear views of the sky, aiming to maximize the number of recorded meteors within common observing fields. A noteworthy trend in recent years has been the rise of pro-am collaborations in this field, involving professional scientists, amateur astronomers, and citizen scientists working together to collect valuable data. These collaborations have significantly expanded the reach of meteor networks, enabling the recording of events from diverse locations and perspectives.

However, with the increasing number of detection stations, the accumulation of video and image data has also surged. Consequently, this data influx has created a bottleneck in processing and analysis, as traditional manual methods prove to be excessively time-consuming and resource-intensive. To deal with these issues, many networks are embracing automation as a means to efficiently handle the significant volumes of generated data \citep{Molau2001ESASP, Spurny2007IAUS, Gural2009JIMO, Brown2010JIMO, Gural2011pimo, Weryk2013Icar, Howie2017ExA, Suk2017EMP, Nikolic2019, Eloy2021MNRAS, Eloy2021Astrodyn, Vida2021MNRAS}. These automated approaches allow meteor scientists to analyze and interpret meteor data faster and more efficiently than ever before, helping to uncover new insights into meteor behavior and properties.

The detection of luminous sources moving in the sky is relatively easy to solve as the camera control software only needs to store and overwrite the last few minutes of recording, and in the event of a sudden increase in illumination, permanently save this data. However, the trigger threshold must be carefully calibrated to avoid missing any meteors while minimizing the number of false positives. Defining this cut-off is complex as it can vary depending on a number of factors, including general lighting conditions, which need to be updated periodically to consider specific dusk and dawn illumination conditions or the presence of the Moon.

The pipelines that attempt to automate the detection and tracking of meteors face a difficult task because meteors are virtually random phenomena and can occur in a variety of ways due to impact geometry, variable velocity, size, shape, composition, viewing angle, sky conditions (clouds or illumination), etc. In addition, meteors must be distinguished from false positives caused by satellites, airplanes, helicopters, drones, birds, lightning, or artificial light sources. The combination of possible characteristics that meteors can exhibit makes it difficult to define fixed parameters that work in all cases. As a result, many networks still rely on human experts to manually review the footage and identify/classify meteors. However, human operators can occasionally make errors, particularly when artificial events cause confusion or ambiguity. However, there are also networks that use fully automated approaches based on traditional computer vision techniques, such as image processing algorithms with fixed instructions, e.g. CAMS \citep{Jenniskens2011Icar}, SonotaCo \citep{SonotaCo2016JIMO}, or EDMOND \citep{Kornos2014pim3}. Some of the detection pipelines currently in use are \textit{MetRec}, \textit{MeteorScan}, and \textit{UFOCapture}; an overview of their capabilities is given in \citet{Molau2005JIMO}. These automated approaches show a high percentage of events with suspicious calculated results due to their reliance on fixed parameters that may not be appropriate for all scenarios \citep{Hajdukova2020PSS}.

Consequently, addressing the challenges of meteor monitoring requires the adoption of artificial intelligence techniques. In this paper, we delve into the utilization of new methodologies for meteor classification and processing tasks. Specifically, we investigate how Machine Learning (ML) approaches can effectively enhance the accuracy and efficiency of automated pipelines, given the massive volume of data generated by meteor networks. We present a fully automated pipeline leveraging Convolutional Neural Networks (CNNs) for meteor detection and tracking using transfer learning and the Gradient-weighted Class Activation Mapping (Grad-CAM) technique.

\section{Artificial intelligence for meteor detection}

Advances in computer technology and hardware performance have fueled the remarkable progress of ML, particularly artificial neural networks with multiple layers, which are classified as deep learning. Neural networks have become increasingly popular in various domains due to their exceptional performance in image classification and recognition. Among them, CNNs have gained popularity for their fault tolerance and self-learning capabilities through multi-layer feedforward networks with a convoluted structure \citep{Gu2018PatRe}. They can handle complex environments and unclear background problems with significantly better generalization ability compared to other methods. A typical CNN architecture includes an input layer, multiple convolutional layers, pooling layers, a fully connected layer, and an output layer. CNNs can be used for both supervised and unsupervised learning, and are applied in diverse fields such as computer vision, natural language processing, and others \citep{Hastie2001}.

In the context of meteor monitoring, current research efforts using ML focus on two main objectives. First, some studies concentrate on determining the presence of a meteor in a given event, with the goal of distinguishing meteoric events from non-meteoric phenomena. Alternatively, other works rely on ML for meteor tracking, facilitating the accurate localization and monitoring of meteoroids throughout their bright atmospheric trajectory. The primary challenge is to effectively distinguish false positives caused by non-meteor objects such as airplanes, birds, and insects, or atmospheric conditions (e.g. clouds). These innovative approaches are being increasingly employed within meteor networks worldwide, including the Global Meteor Network (GMN) \citep{Gural2019}, AllSky7 Fireball Network Germany (FNG\footnote{\url{https://allsky7.net/}}), Meteorite Orbits Reconstruction by Optical Imaging (MOROI) \citep{Nedelcu18moroi}, Canadian Automated Meteor Observatory (CAMO) \citep{Weryk13camo}, Cameras for All-sky Meteor Surveillance (CAMS) \citep{JENNISKENS2011cams}, EXOSS meteor network\footnote{\url{https://exoss.org/}}, and Meteor Automatic Imager and Analyzer (MAIA) \citep{Vitek11maia}.

In image classification, it is customary to use transfer learning techniques with pre-trained models of CNNs \citep{Sennlaub22, Marsola19, galindo18}. These methods allow inheriting the ability to detect objects from those pre-trained models, which need to be re-trained on meteors. With this methodology, optimal results are achieved with a smaller number of training data compared to starting from an uninitialized model. As underlined in \citet{galindo18}, uttermost results are achieved when the proper pre-trained dataset is selected. They compared the performance of ImageNET and Fashion-MNIST \citep{xiao2017} with fine-tuning, concluding that the latter is the most optimal as it is already trained to work with black and white images. They also checked whether the CNN could distinguish meteors if the image was previously tweaked (slightly zoomed, rotated, or flipped). The results showed that these transformations could produce unrealistic apparent trajectories and worsen the classification. In order to solve these types of problems, \citet{Ganju2023} use a windowing technique to create new frames from existing ones. With this, all meteor detection would have the same number of frames, easing the posterior analysis. In \citet{Cecil20} is shown a comparison of different sets of combinations of different techniques used in image processing, such as convolutions and max-pools. 

Even though most CNN meteor detection algorithms have been trained to reach a satisfying prediction percentage ($>99\%$), particularly when considering large sample sizes of more than $\sim$10,000 events, some anomalies are still misclassified as meteors for small datasets. The next step in meteor detection algorithms is to consider the intrinsic properties of meteors on camera images to discard these misleading anomalies. Additionally, the main drawback of the current meteor tracking algorithms is the runtime required to analyze high-definition 1080p video images. However, they perform well when dealing with small, low-resolution video images.

Beyond CNNs, other ML techniques are often used. It is the case of Recurrent Neural Networks (RNN), Gradient Boost (GB), or Random Forest (RF), which can also be used as complementary analysis tools \citep{Gural2019, Anghel22}. Temporal resolution can be introduced in the analysis by using other networks such as Long Short-Term Memory (LSTM), Gated Recurrent Unit (GRU), Temporal Convolutional Network (TCN), Time Delay Neural Network (TDNN), Support Vector Machine (SVM), or VGG16 \citep{Siladji2015moconf, simonyan15vgg, Sennlaub22}, supporting 16 layers.  A particular additional technique is the discrete pulse transform (DPT) \citep{RohwerLaurie06}, in which image signals are decomposed in pulses. \citet{VitekNasyrova19} introduced DPT to characterize the number of pulses related to the meteor compared to those of stars. Even though it is not specified if the results are better than other works, they do underline that using DPT is faster than other methods used in the MAIA data.

\citet{Sennlaub22} also classified those false positives based on their origin. They did not achieve solid statements but pointed out similarities among false positive subgroups. As sketched in \citet{LeLan21} a future algorithm to classify the false positives would include prior knowledge of each subgroup. These could also be expanded to classify the re-entry of artificial space debris. They could also include a cross-matching identification between different stations as a method of improving the overall accuracy \citep{Anghel23}.

Table \ref{tab:review} provides a comprehensive summary of the outcomes achieved in these works accompanied by the data source, the number of samples, the technique used, and the results such as F1 score and accuracy. The F1 score is a performance metric that combines precision and recall to measure the accuracy of a classification model. Precision represents the proportion of true positive predictions out of all positive predictions, while recall represents the proportion of true positive predictions out of all actual positive instances. The F1 score considers both precision and recall, making it useful when the dataset is imbalanced or when false positives and false negatives have different consequences. Accuracy is a common evaluation metric used to measure the overall correctness of a classification model. It calculates the proportion of correctly predicted instances (both true positives and true negatives) out of the total number of instances in the dataset.

\begin{table*}[tbp]
\caption{Summary of the main previous works on ML applied to meteor detection. For each contribution, we denote the meteor network source, the number of samples used to train the models, their goal, the transfer learning method used (if any), the different models compared, and the best F1 score and/or accuracy achieved (when provided). F1 score and accuracy have been normalized between 0 and 1.}
\label{tab:review}
\resizebox{\textwidth}{!}{%
\begin{tabular}{lccccccc}
\hline
Article & Meteor network & \# samples & Goal & Transfer learning & Model & F1 score & Accuracy \\
\hline
\citet{Ganju2023} & CAMS & 19,152 & Detection & BiLSTM & - & 0.91 & 0.89 \\
\citet{Anghel23} & MOROI &  9,858 & Detection & - & CNN, GB and RF & 0.989 & 0.998 \\
\citet{Sennlaub22} & AllSky7 FNG & 20,000 & Detection & ImageNET & CNN: GRU and SVM & - & 0.991 \\
\citet{Cecil20} & CAMO & 50,745 & Tracking & No & CNN: multi-layer and max pool & 0.997 & 0.998 \\
\citet{Gural2019} & CAMS & 200,361 & Tracking & MeteorNet & RNN and LSTM & 0.974 & 0.981 \\
\citet{Marsola19} & EXOSS & 400 & Detection & ImageNET & VGG16 & - & 0.8435 \\
\citet{VitekNasyrova19} & MAIA & - & Tracking & No & DPT & - & -\\
\citet{galindo18} & EXOSS & 1,660 & Detection & Fashion-MNIST & - & 0.94 & 0.960 \\
\hline
\end{tabular}%
}
\end{table*}

\section{CNN, transfer learning, and Grad-CAM}

Our work aims to use ML techniques to achieve two main goals: detecting the presence of meteors in images and tracking the motion of meteors in the field of view. To achieve this, and based on the review of the scientific literature, we have decided to develop a CNN model that classifies images into two groups, "Meteors" and "No-meteors", using transfer learning. In addition, we implemented a novel application of Grad-CAM to track the coordinates of the meteor's motion.

\subsection{Detection}

\begin{figure}[tbp]
\begin{center}
\includegraphics[width=0.8\linewidth]{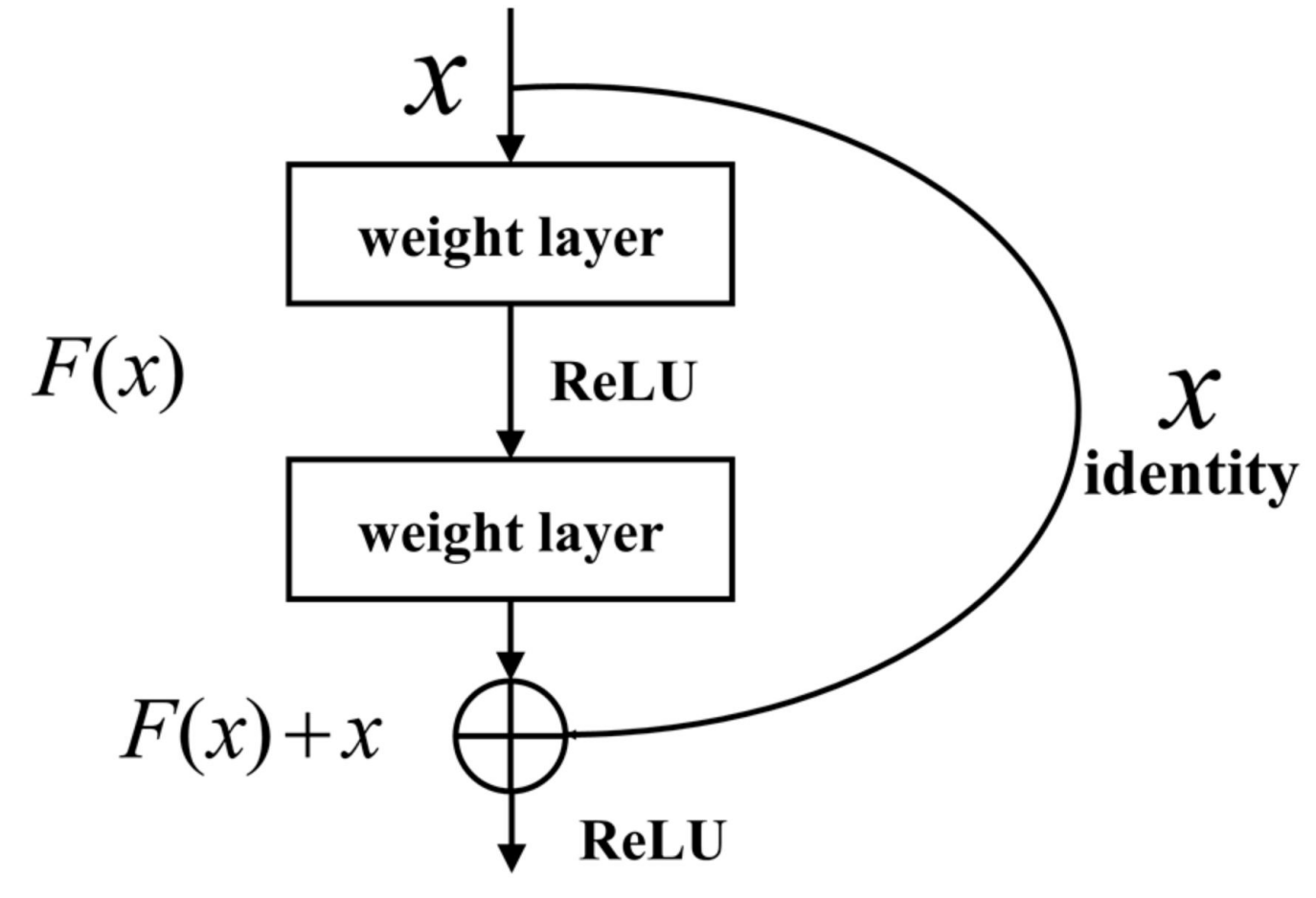}
\caption{A basic-block building block of ResNet-34.
\label{fig:basic-block}}
\end{center}
\end{figure}

To build our model, we chose to use ResNet-34, which is a 34-layer pre-trained CNN from the Residual Network family \citep{He2015arXiv}. This allowed us to quickly specialize the model for our specific use case using transfer learning techniques with a small dataset and rapidly inheriting all detection skills already learned by the network. ResNet-34 mainly consists of an input layer, convolutional layers, residual blocks, shortcut connections, downsampling, global average pooling, and fully connected layers. One of the key elements of this network is the residual building block, which is its infrastructure. As shown in Figure \ref{fig:basic-block}, the residual building block consists of several convolutional layers (Conv), batch normalizations (BN), a rectified linear unit (ReLU) activation function, and a link. This block is used for all 34 layers of ResNet-34, as depicted in Figure \ref{tab:ResNet34_structure}. The output of the residual block is given by the formula $y = F(x) + x$, where $F$ is the residual function and $x$ and $y$ are the input and output of the residual function, respectively. The entire residual network is composed of the first convolutional layer and multiple basic blocks, making it a highly effective and efficient deep learning architecture for image recognition tasks.

\begin{table}[tbp]
\centering
\footnotesize
\caption{The structure of ResNet-34.}
\label{tab:ResNet34_structure}
\begin{tabular}{ccc}
\hline
Layer name & Output size & 34-layer \\
\hline
Conv1 & 112$\times$112 & $\begin{matrix}7\times7, 64, stride 2\\ 3\times3\,\,max\,\,pool, stride 2 \end{matrix}$ \\
Conv2\_x & 56$\times$56 & $\begin{bmatrix} 3\times3, 64\\ 3\times3, 64 \end{bmatrix}\times3$  \\
Conv3\_x & 28$\times$28 & $\begin{bmatrix} 3\times3, 128\\ 3\times3, 128 \end{bmatrix}\times4$  \\
Conv4\_x & 14$\times$14 & $\begin{bmatrix} 3\times3, 256\\ 3\times3, 256 \end{bmatrix}\times6$ \\
Conv5\_x & 7$\times$7 & $\begin{bmatrix} 3\times3, 512\\ 3\times3, 512 \end{bmatrix}\times3$ \\
 & 1$\times$1 & average pool, 1000-d fc, softmax \\
\hline
\end{tabular}
\end{table}

For training and testing the model, we used a dataset of 982 images of meteors detected by optical devices of the Spanish Meteor Network (SPMN) network stations \citep{Trigo2006AG}, along with 56,285 images without meteor detection collected over the year 2021, particularly from the Pujalt observatory. To balance the two groups, we generated a dataset of 982 images with meteors and 1,050 without detections for training. To ensure reliable model performance, a portion of these images, specifically 20\%, is allocated for validation, which is utilized to evaluate the training progress. From this dataset, 300 images were specifically set aside for testing purposes, 150 from the meteor class and 150 from the no-meteor. The test set serves as an independent dataset to evaluate the final performance of the trained model after the completion of the training process.

The dataset consisted of grayscale long exposure (30 seconds) images that were pre-processed to enhance the meteor trail and remove static visual elements from the background by subtracting consecutive images. This included converting the images to black and white, resizing them to 400x400, and subtracting successive images to remove the background. To facilitate the generalization of the model and reduce overfitting, data augmentation techniques were used during the transfer learning process. Specifically, each batch of images received by the CNN during the 35 epochs of training was modified with geometric transformations such as randomly flipping, cropping, rotating, and translating the images, or applying lighting modifications.

\subsection{Tracking}

The final layer of our CNN exhibits activations corresponding to neurons that are specifically triggered when a meteor is detected in an image. While such activations provide initial utility, leveraging the subsequent classification layer's weights in conjunction with these activations further enhances their significance. It is important to note that the classification layer possesses a comprehensive understanding of the meteor classification task, enabling it to assign appropriate weights to the activated neurons. Hence, certain previously activated neurons may be deemed less influential in the final classification decision. This is the foundation of the so-called Class Activation Mapping (CAM) technique \citep{Zhou2015arXiv151204150Z}.

\begin{figure}[tbp]
\includegraphics[width=\linewidth]{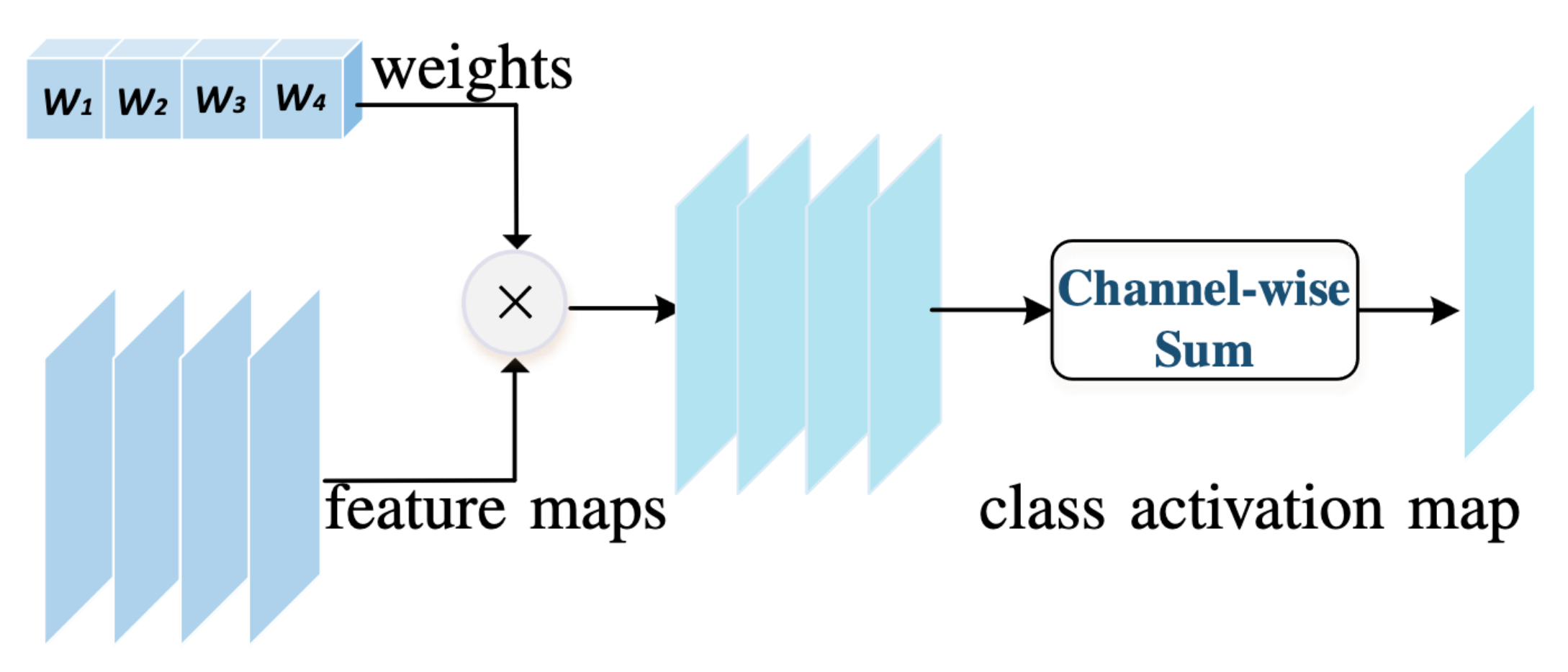}
\caption{The general process of class activation mapping method. Adapted from \citet{Jiang2021ITIP}.
\label{fig:CAM_method}}
\end{figure}

The CAM technique is usually employed to generate a heatmap for a given class (class "meteor" in our case). This technique involves capturing the activations from the last convolutional layer of the CNN and multiplying them by the corresponding weights from the last fully connected layer responsible for the classification task. By performing this operation, the CAM technique effectively highlights the areas within the image that contribute the most to the meteor classification, that is, the Region of Interest (ROI). Figure \ref{fig:CAM_method} illustrates the overall procedures of this method.

However, in order to further enhance the performance of our model, we opted to incorporate an advanced variant of CAM known as Grad-CAM \citep{Selvaraju2016}. Grad-CAM builds upon the CAM methodology by integrating gradient information instead of the weights of the classification layer. This provides a more fine-grained localization of important regions within an image. Grad-CAM computes the gradients of the target class of a specific layer with respect to the activations of the same layer. By multiplying the activations from the last convolutional layer with the average of the gradients across the feature map of that layer, Grad-CAM obtains the importance weights for the activation maps. The weighted combination of the activation maps produces the final heatmap, which visually highlights the critical regions within the input image for the classification of the target class.

\begin{sidewaysfigure*}
\includegraphics[width=1\textwidth]{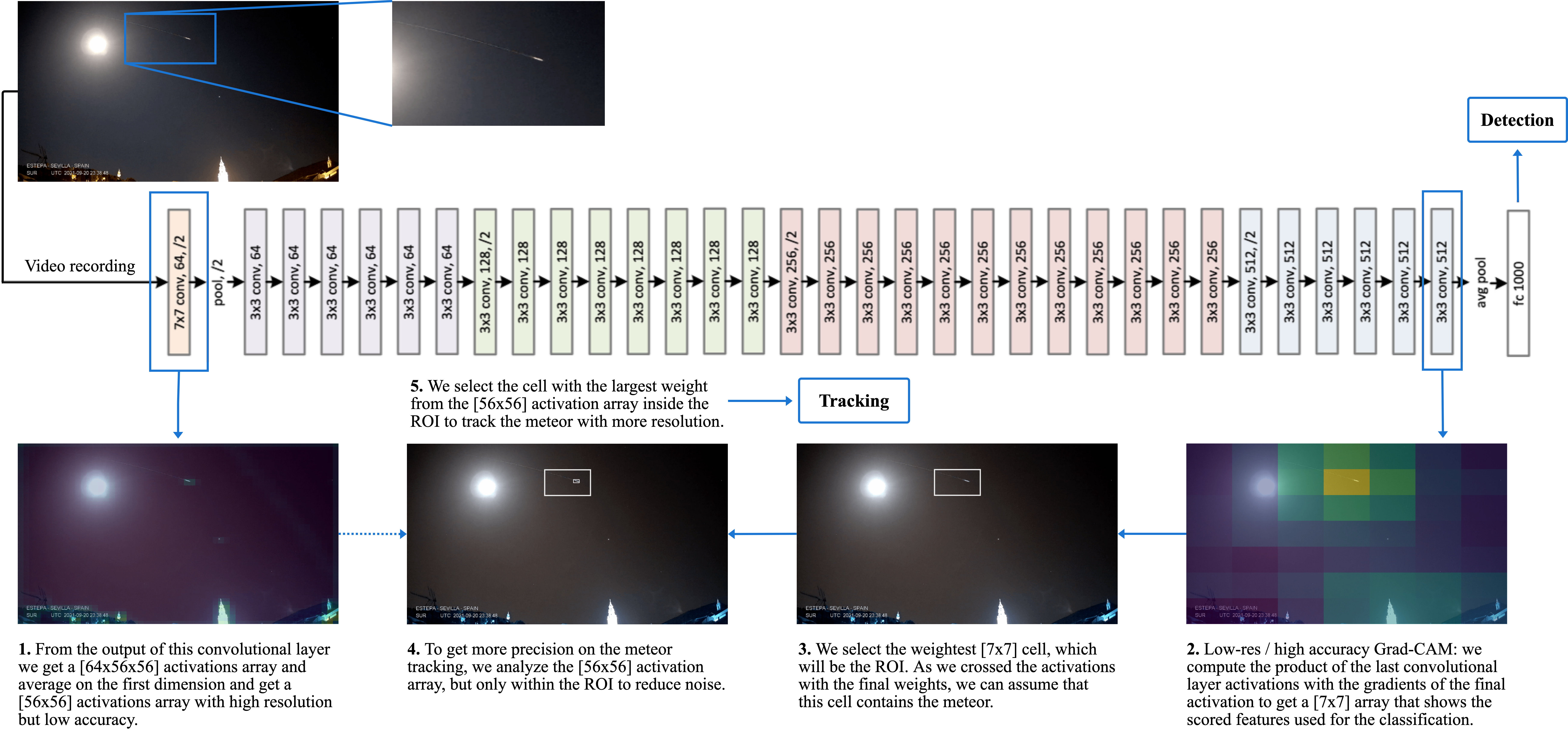}
\caption{Illustrative diagram of the proposed process of meteor detection and tracking using CNN and Grad-CAM. The image exemplifies a video detection made by the SPMN station in Estepa, Seville operated by Antonio J. Robles.
\label{fig:diagram}}
\end{sidewaysfigure*}

To capture finer details for properly tracking the meteor motion, we focus on the activations of the initial convolutional layer, restricting our attention to activations falling within the defined ROI delineated by the Grad-CAM. Within the CNN, the activation of the deepest layers offers a more detailed resolution map (56x56). However, these layers are more prone to noise and less reliable in accurately identifying the ROI related to meteor detection. To tackle these difficulties, we apply a noise reduction strategy by selectively retaining only the cells that align with the cells from the higher precision but lower resolution Grad-CAM (7x7) with a non-zero value. By doing so, we filter out noisy activations and focus on the cells that have a meaningful impact on the meteor classification. Subsequently, we extract the cells with the maximum activation values from the refined high-resolution activation map. By calculating the average position among these selected cells, we are able to project a single point onto the original image. This refinement process significantly enhances the accuracy of meteor detection by precisely pinpointing the location of meteors within the frames generated by our model. Figure \ref{fig:diagram} illustrates the described meteor detection and tracking process.

Note that we do not multiply the activations of the initial layer with the weights of the classification layer because the spatial and semantic gap between the initial convolutional layer and the classification layer limits the effectiveness of such an approach. Although Grad-CAM allows the analysis of any layer in the network, we observed that focusing just on the activations from the initial convolutional layer within the ROI calculated using Grad-CAM on the last layer yields good results.

\section{Training and results}

\begin{figure}[ht!]
\includegraphics[width=\linewidth]{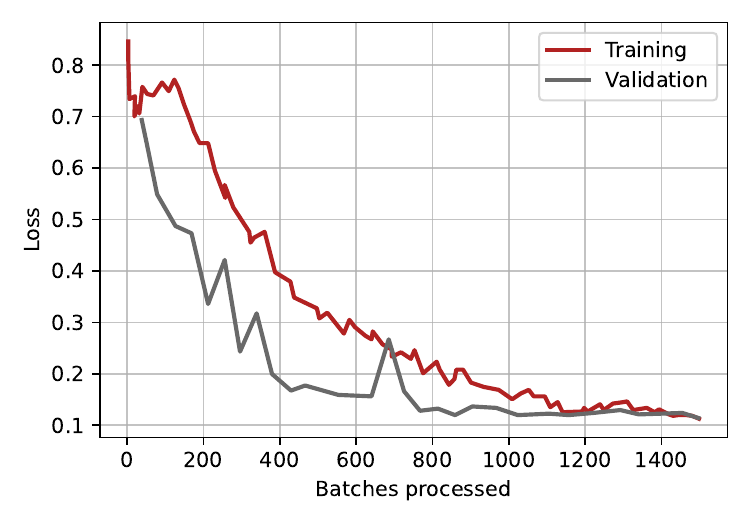}
\caption{Training and validation loss during model training.
\label{fig:train_valid}}
\end{figure}

Figure \ref{fig:train_valid} shows the evolution during the training process of the loss function as a function of the batches processed, which is a metric that measures the deviation between predicted and actual values. The primary goal during training is to minimize this function. The training set consists of the images used to train the CNN model, while the validation set consists of a subset of 150 images randomly selected and reserved for evaluating the model's performance and generalization ability. By evaluating the loss function on both sets, we monitor the progress of the model and detect any signs of overfitting or underfitting. 

A batch refers to a group of images that are processed together during each iteration of the training algorithm. The total number of batches processed can be calculated using the formula $Batches = N * (I / BS)$, where $N$ is the number of epochs, which is the number of times the entire training data set is passed through the network during training. $I$ denotes the total number of images in the training dataset, including both meteor and non-meteor images. Finally, $BS$ refers to the batch size, which represents the number of images fed to the network in each training iteration. For this study, we used a batch size of 32. 

By substituting these values, we determined that a total of 1,515 batches were processed during the training process. This corresponds to the maximum value observed in Figure \ref{fig:train_valid}, indicating the completion of all batches. Understanding the relationship between the loss function, the batches processed, and the progress of the training and validation sets provides valuable insight into the training dynamics and performance evaluation of the neural network model. As every batch comprises 32 images and is subjected to various random transformations, the overall training process, including data augmentation, incorporates a varied array of 48,480 distinct images. Table \ref{tab:training} compiles the evolution of different metrics during the training process.

\begin{table}[ht!]
\centering
\footnotesize
\caption{Metrics evolution during the training process, including for each epoch the error rate, accuracy, precision, recall, and F1 score.}
\label{tab:training}
\begin{tabular}{ccccc}
\hline
Epoch & Error & Accuracy & Precision & F1 \\
\hline
0 & 0.245665 & 0.754335 & 0.763323 & 0.753592 \\
1 & 0.210983 & 0.789017 & 0.788704 & 0.788804 \\
2 & 0.205202 & 0.794798 & 0.798389 & 0.794714 \\
3 & 0.187861 & 0.812139 & 0.823898 & 0.811444 \\
4 & 0.127168 & 0.872832 & 0.872594 & 0.872726 \\
5 & 0.161850 & 0.838150 & 0.842479 & 0.838064 \\
6 & 0.104046 & 0.895954 & 0.895722 & 0.895867 \\
7 & 0.138728 & 0.861272 & 0.875206 & 0.860709 \\
8 & 0.086705 & 0.913295 & 0.913068 & 0.913222 \\
9 & 0.072254 & 0.927746 & 0.928404 & 0.927527 \\
10 & 0.063584 & 0.936416 & 0.936314 & 0.936382 \\
11 & 0.049133 & 0.950867 & 0.950896 & 0.950775 \\
12 & 0.060694 & 0.939306 & 0.939599 & 0.939294 \\
13 & 0.063584 & 0.936416 & 0.936896 & 0.936244 \\
14 & 0.049133 & 0.950867 & 0.950896 & 0.950775 \\
15 & 0.095376 & 0.904624 & 0.912476 & 0.904489 \\
16 & 0.066474 & 0.933526 & 0.936153 & 0.933521 \\
17 & 0.049133 & 0.950867 & 0.950693 & 0.950834 \\
18 & 0.052023 & 0.947977 & 0.948137 & 0.947865 \\
19 & 0.037572 & 0.962428 & 0.962805 & 0.962337 \\
20 & 0.052023 & 0.947977 & 0.948137 & 0.947865 \\
21 & 0.043353 & 0.956647 & 0.956495 & 0.956586 \\
22 & 0.052023 & 0.947977 & 0.948109 & 0.947961 \\
23 & 0.046243 & 0.953757 & 0.953543 & 0.953719 \\
24 & 0.037572 & 0.962428 & 0.962282 & 0.962375 \\
25 & 0.037572 & 0.962428 & 0.962805 & 0.962337 \\
26 & 0.049133 & 0.950867 & 0.950896 & 0.950775 \\
27 & 0.034682 & 0.965318 & 0.965915 & 0.965224 \\
28 & 0.043353 & 0.956647 & 0.956688 & 0.956566 \\
29 & 0.049133 & 0.950867 & 0.950693 & 0.950834 \\
30 & 0.049133 & 0.950867 & 0.950693 & 0.950834 \\
31 & 0.052023 & 0.947977 & 0.947761 & 0.947933 \\
32 & 0.046243 & 0.953757 & 0.953551 & 0.953702 \\
33 & 0.046243 & 0.953757 & 0.953551 & 0.953702 \\
34 & 0.034682 & 0.965318 & 0.965525 & 0.965244 \\
\hline
\end{tabular}
\end{table}

The next step involves defining the crucial hyperparameter for the training process: the learning rate. By evaluating the loss function values across various learning rates, we can identify the region of sustained and substantial loss reduction, disregarding transient peaks and irregular drops as they do not represent reliable trends. Once this region is identified, we pinpoint the midpoint of the steepest descent line, which corresponds to the most significant loss reduction. This specific learning rate is then selected for the subsequent training procedure. For our specific study, we determined a learning rate of 0.003 to be the optimal choice.

The training phase of our pipeline ends with an F1 score of 0.94, indicating the high precision and recall achieved during the training process. We then evaluated the performance of the trained model on the test dataset to assess its generalization capabilities. The evaluation results show that our model achieved an accuracy of 0.96 on the test dataset. This accuracy metric indicates the model's ability to correctly classify meteor and non-meteor images with a high degree of accuracy. Furthermore, when specifically considering the meteor class, our model achieved a precision of 0.98.

\begin{figure}[ht!]
\includegraphics[width=\linewidth]{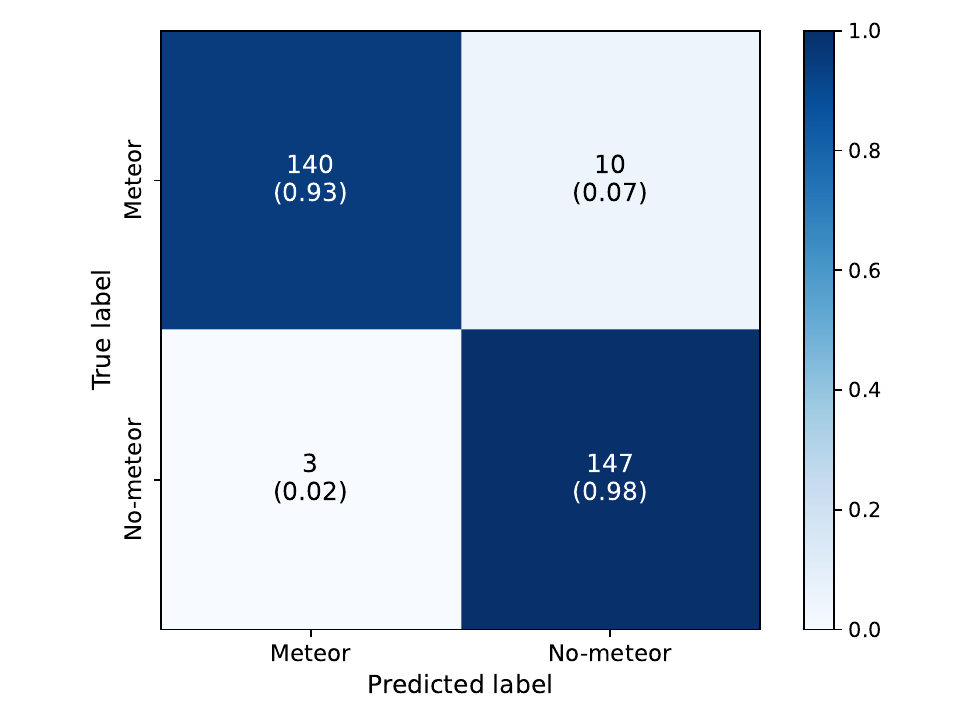}
\caption{Confusion matrix of the trained model with normalized values in parentheses. 
\label{fig:confusion}}
\end{figure}

To further analyze the performance of the model, a confusion matrix was constructed as it provides insight into the classification performance by showing the distribution of the predicted labels against the true labels. The confusion matrix for our two labels "Meteor" and "No-Meteor" is shown in Figure \ref{fig:confusion}.

In the confusion matrix, the rows correspond to the true labels, while the columns represent the predicted labels. The matrix values indicate the proportion of images belonging to each category. The confusion matrix shows that 47\% of the images were correctly classified as meteors, while 3.3\% of the images were incorrectly classified as non-meteor images. Furthermore, 1\% of the images were incorrectly classified as meteor images, while 49\% were correctly identified as non-meteor images. The high accuracy achieved by the pipeline demonstrates its robustness in accurately detecting and classifying meteor images. The low misclassification rates for both meteor and non-meteor classes indicate the effectiveness of the trained CNNs in distinguishing between these classes, then minimizing the required human time for these time-consuming tasks. 

It is worth noting that the proposed model shows the ability to generalize and accurately detect meteor trails even in color frames that have not undergone the previous frame subtraction process. This is particularly noteworthy because these frames may contain stationary elements such as clouds, the Moon, buildings, or other obstructions that can cause interference and affect the accuracy of meteor detection. Despite these challenges, the model can effectively distinguish and identify meteor trails, providing robust and reliable results. This further validates the model's ability to operate under real-world conditions, making it a valuable tool for meteor scientists and enthusiasts alike.

\begin{figure}[ht!]
\includegraphics[width=\linewidth]{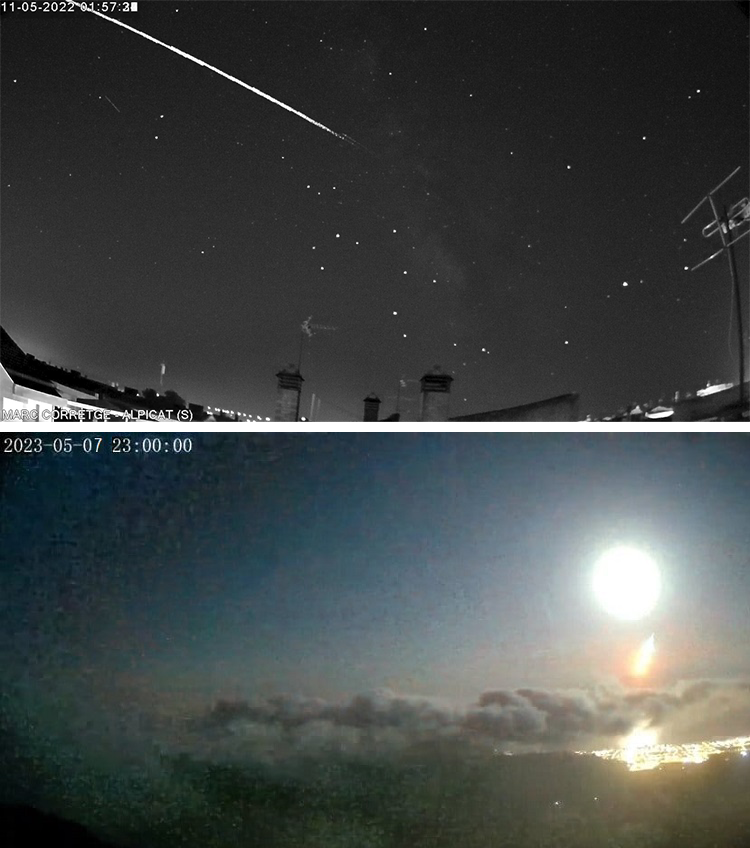}
\caption{Top panel: False positive of a \textit{SpaceX Starlink} satellite track as it exhibits similar characteristics as a meteor trail. Recording obtained from the Alpicat SPMN station under the operation of Marc Corretgé-Gilart. Bottom panel: False negative of SPMN070523G superbolide recorded near the full moon with a cloudy sky. Recording obtained from Bartolo-Castelló SPMN station under the operation of Vicente Ibañez}. 
\label{fig:wrongdet}
\end{figure}

However, this very capability in detecting meteor trails renders it susceptible to satellite misidentifications, as they often exhibit a similar trace when reflecting the sunlight. An instance of such incorrect classification is illustrated in the top panel of Figure \ref{fig:wrongdet} with a transit of a \textit{SpaceX Starlink} satellite train, necessitating retraining the network to encompass this new class. Conversely, the bottom panel shows an undetected meteor, possibly due to its proximity to the full moon and dimming by a cloud-covered field of view.

We used data augmentation techniques during the training process, which is a common strategy that helps prevent overfitting and improves the model's ability to generalize well to unseen data. However, it is worth noting that \citet{xiao2017} suggested data augmentation could potentially degrade model performance in certain cases. Therefore, it is possible that our use of data augmentation in the training process may have resulted in slightly lower performance compared to some of the models reported in Table \ref{tab:review}. Despite this slight performance difference, our pipeline still demonstrates a high level of accuracy and efficiency in meteor detection and classification. The inclusion of data augmentation techniques is crucial to promote better generalization and robustness in the model, even though it may have had a slight impact on the overall results compared to other models in the comparison.

We compared our results with those obtained by \citet{galindo18} and \citet{Marsola19}, who used datasets of similar size to ours. In this comparison, our methodology equals or outperforms existing approaches, demonstrating superior performance. Furthermore, our pipeline incorporates the added complexity of meteor tracking, which presents a significant challenge due to the smaller luminous trace in each frame. Meteor tracking enables the computation of the velocity curve, a key factor in both discriminating between artificial and natural objects and in determining the heliocentric orbit and the potential meteorite strewn field. Automating this process facilitates the extraction of orbital elements for hundreds or thousands of meteors detected nightly, providing valuable insights into both the sporadic meteoroid background flux and the characteristics of meteoroid streams.

In subsequent phases, we intend to enhance the pipeline by refining the balance of false positives, encompassing a diverse spectrum of potential false positive sources including satellites, planes, birds, bugs, and other light sources in the training process.

\section{Conclusions}

The implementation of automated detection software has led to a massive increase in the amount of data collected and reduced every year by meteor networks. However, the need for human oversight to filter out false positives and organize the records has created a bottleneck, and the traditional computer vision techniques implemented have limited performance due to the random and specific characteristics of each meteor event. We employed CNNs to address these challenges. 

In our study, we used a dataset of 982 meteor images along with 1,050 images without meteors detected by SPMN stations in 2021 to train a CNN model. A transfer learning technique was applied, and Grad-CAM was used for accurate tracking. Our main results are as follows:

1) Our approach utilized ResNet-34, a deep learning architecture consisting of 34 pre-trained layers. By using pre-trained layers, we capitalized on the knowledge and representations gained from a large dataset during the initial training phase, resulting in improved model performance. In addition, data augmentation techniques were employed to facilitate the model's ability to accurately generalize and mitigate overfitting. The results achieved demonstrate a precision of 98\% for meteor classification.

2) Grad-CAM was used to track the coordinates of the meteor within each image. This technique involved analyzing deeper layers of the neural network, which have higher accuracy but lower resolution. ROI information was extracted using the gradients of the last convolutional layer and then combined with activation information from the initial layer, which is characterized by higher resolution but lower accuracy. This fusion of information allowed the identification of the most critical pixel corresponding to the position of the meteoroid. This technique allows for the precise localization of meteor positions within frames.

3) The high performance achieved by our pipeline underscores its robustness in precisely detecting and classifying meteor images. The success rate, even with a relatively small dataset, highlights the potential of our method to significantly reduce the workload of meteor scientists and station operators involved in meteor data analysis. This potential is further enhanced by one of the most notable advancements in our methodology: the novel use of Grad-CAM for meteor tracking in combination with the initial activation map.

In summary, our study highlights the significant potential of applying ML techniques to meteor monitoring. It illustrates the effectiveness of CNNs and transfer learning in reducing false positives and correctly identifying meteors in images. By automating the meteor monitoring process, our pipeline increases the efficiency of meteor detection and tracking using the Grad-CAM technique. This, in turn, facilitates the study of meteoroid fluxes, aids in population characterization, and improves our ability to distinguish between meteorite-dropping events, thereby increasing fresh extraterrestrial material recovery rates.

    \section*{Acknowledgments}
      This project has received funding from the European Research Council (ERC) under the European Union’s Horizon 2020 research and innovation programme (grant agreement No. 865657) for the project “Quantum Chemistry on Interstellar Grains” (QUANTUMGRAIN). JMT-R, EP-A, and PG-T acknowledge financial support from the project PID2021-128062NB-I00 funded by MCIN/AEI/10.13039/501100011033. AR acknowledges financial support from the FEDER/Ministerio de Ciencia e Innovación – Agencia Estatal de Investigación (PID2021-126427NB-I00, PI: AR). This work was also partially supported by the program Unidad de Excelencia María de Maeztu CEX2020-001058-M. We thank all SPMN station operators and photographers whose continuous dedication has allowed to record incessantly the meteors of the Iberian Peninsula, the Balearic Islands, and the Canary Islands. We express our gratitude to David Puiggròs-Figueras for his assistance in the CAM and Grad-CAM methodology. The following \textit{Python} packages were extensively used in the study: \textit{Numpy}, \textit{OpenCV}, \textit{Matplotlib}, \textit{PyTorch}, and \textit{fast.ai}.

\bibliographystyle{elsarticle-harv} 
\bibliography{cas-refs}

\begin{thebibliography}{64}
\expandafter\ifx\csname natexlab\endcsname\relax\def\natexlab#1{#1}\fi
\providecommand{\url}[1]{\texttt{#1}}
\providecommand{\href}[2]{#2}
\providecommand{\path}[1]{#1}
\providecommand{\DOIprefix}{doi:}
\providecommand{\ArXivprefix}{arXiv:}
\providecommand{\URLprefix}{URL: }
\providecommand{\Pubmedprefix}{pmid:}
\providecommand{\doi}[1]{\href{http://dx.doi.org/#1}{\path{#1}}}
\providecommand{\Pubmed}[1]{\href{pmid:#1}{\path{#1}}}
\providecommand{\bibinfo}[2]{#2}
\ifx\xfnm\relax \def\xfnm[#1]{\unskip,\space#1}\fi
\bibitem[{{Anghel} et~al.(2022){Anghel}, {Nedelcu}, {Birlan} and
  {Boaca}}]{Anghel22}
\bibinfo{author}{{Anghel}, S.}, \bibinfo{author}{{Nedelcu}, D.A.},
  \bibinfo{author}{{Birlan}, M.}, \bibinfo{author}{{Boaca}, I.},
  \bibinfo{year}{2022}.
\newblock \bibinfo{title}{{Machine learning methods applied to meteor detection
  filtering}}, in: \bibinfo{booktitle}{European Planetary Science Congress},
  pp. \bibinfo{pages}{EPSC2022--1107}.
\newblock \DOIprefix\doi{10.5194/epsc2022-1107}.
\bibitem[{{Anghel} et~al.(2023){Anghel}, {Nedelcu}, {Birlan} and
  {Boaca}}]{Anghel23}
\bibinfo{author}{{Anghel}, S.}, \bibinfo{author}{{Nedelcu}, D.A.},
  \bibinfo{author}{{Birlan}, M.}, \bibinfo{author}{{Boaca}, I.},
  \bibinfo{year}{2023}.
\newblock \bibinfo{title}{{Single-station meteor detection filtering using
  machine learning on MOROI data}}.
\newblock \bibinfo{journal}{Monthly Notices of the Royal Astronomical Society}
  \bibinfo{volume}{518}, \bibinfo{pages}{2810--2824}.
\newblock \DOIprefix\doi{10.1093/mnras/stac3229}.
\bibitem[{Bottke et~al.(2002)Bottke, Cellino, Paolicchi and
  Binzel}]{Bottke2002}
\bibinfo{editor}{Bottke, W.F.}, \bibinfo{editor}{Cellino, A.},
  \bibinfo{editor}{Paolicchi, P.}, \bibinfo{editor}{Binzel, R.P.} (Eds.),
  \bibinfo{year}{2002}.
\newblock \bibinfo{title}{Asteroids {III}}.
\newblock \bibinfo{publisher}{University of Arizona Press}.
\newblock \URLprefix \url{https://doi.org/10.2307/j.ctv1v7zdn4},
  \DOIprefix\doi{10.2307/j.ctv1v7zdn4}.
\bibitem[{{Brown} et~al.(2010){Brown}, {Weryk}, {Kohut}, {Edwards} and
  {Krzeminski}}]{Brown2010JIMO}
\bibinfo{author}{{Brown}, P.}, \bibinfo{author}{{Weryk}, R.J.},
  \bibinfo{author}{{Kohut}, S.}, \bibinfo{author}{{Edwards}, W.N.},
  \bibinfo{author}{{Krzeminski}, Z.}, \bibinfo{year}{2010}.
\newblock \bibinfo{title}{{Development of an All-Sky Video Meteor Network in
  Southern Ontario, Canada The ASGARD System}}.
\newblock \bibinfo{journal}{WGN, Journal of the International Meteor
  Organization} \bibinfo{volume}{38}, \bibinfo{pages}{25--30}.
\bibitem[{{Cecil} and {Campbell-Brown}(2020)}]{Cecil20}
\bibinfo{author}{{Cecil}, D.}, \bibinfo{author}{{Campbell-Brown}, M.},
  \bibinfo{year}{2020}.
\newblock \bibinfo{title}{{The application of convolutional neural networks to
  the automation of a meteor detection pipeline}}.
\newblock \bibinfo{journal}{Planetary and Space Science} \bibinfo{volume}{186},
  \bibinfo{pages}{104920}.
\newblock \DOIprefix\doi{10.1016/j.pss.2020.104920}.
\bibitem[{{Ceplecha}(1987)}]{Ceplecha1987}
\bibinfo{author}{{Ceplecha}, Z.}, \bibinfo{year}{1987}.
\newblock \bibinfo{title}{{Geometric, Dynamic, Orbital and Photometric Data on
  Meteoroids From Photographic Fireball Networks}}.
\newblock \bibinfo{journal}{Bulletin of the Astronomical Institutes of
  Czechoslovakia} \bibinfo{volume}{38}, \bibinfo{pages}{222}.
\bibitem[{{Ceplecha} et~al.(1998){Ceplecha}, {Borovi{\v{c}}ka}, {Elford},
  {Revelle}, {Hawkes}, {Porub{\v{c}}an} and {{\v{S}}imek}}]{Ceplecha1998SSRv}
\bibinfo{author}{{Ceplecha}, Z.}, \bibinfo{author}{{Borovi{\v{c}}ka}, J.},
  \bibinfo{author}{{Elford}, W.G.}, \bibinfo{author}{{Revelle}, D.O.},
  \bibinfo{author}{{Hawkes}, R.L.}, \bibinfo{author}{{Porub{\v{c}}an}, V.},
  \bibinfo{author}{{{\v{S}}imek}, M.}, \bibinfo{year}{1998}.
\newblock \bibinfo{title}{{Meteor Phenomena and Bodies}}.
\newblock \bibinfo{journal}{Space Science Reviews} \bibinfo{volume}{84},
  \bibinfo{pages}{327--471}.
\newblock \DOIprefix\doi{10.1023/A:1005069928850}.
\bibitem[{{Colas} et~al.(2020){Colas}, {Zanda}, {Bouley}, {Jeanne}, {Malgoyre},
  {Birlan}, {Blanpain}, {Gattacceca}, {Jorda}, {Lecubin} and
  {Zollo}}]{Colas2020AA}
\bibinfo{author}{{Colas}, F.}, \bibinfo{author}{{Zanda}, B.},
  \bibinfo{author}{{Bouley}, S.}, \bibinfo{author}{{Jeanne}, S.},
  \bibinfo{author}{{Malgoyre}, A.}, \bibinfo{author}{{Birlan}, M.},
  \bibinfo{author}{{Blanpain}, C.}, \bibinfo{author}{{Gattacceca}, J.},
  \bibinfo{author}{{Jorda}, L.}, \bibinfo{author}{{Lecubin}, ..., R.},
  \bibinfo{author}{{Zollo}, A.}, \bibinfo{year}{2020}.
\newblock \bibinfo{title}{{FRIPON: a worldwide network to track incoming
  meteoroids}}.
\newblock \bibinfo{journal}{Astronomy and Astrophysics} \bibinfo{volume}{644},
  \bibinfo{pages}{A53}.
\newblock \DOIprefix\doi{10.1051/0004-6361/202038649},
  \href{http://arxiv.org/abs/2012.00616}{{\tt arXiv:2012.00616}}.
\bibitem[{{Drolshagen} et~al.(2021){Drolshagen}, {Ott}, {Koschny},
  {Drolshagen}, {Vaubaillon}, {Colas}, {Zanda}, {Bouley}, {Jeanne}, {Malgoyre},
  {Birlan}, {Vernazza}, {Gardiol}, {Nedelcu}, {Rowe}, {Forcier},
  {Trigo-Rodriguez}, {Pe{\~n}a-Asensio}, {Lamy}, {Ferri{\`e}re}, {Barghini},
  {Carbognani}, {Di Martino}, {Rasetti}, {Valsecchi}, {Volpicelli}, {Di Carlo},
  {Knapic}, {Pratesi}, {Riva}, {Stirpe}, {Zorba}, {Hernandez}, {Grandchamps},
  {Jehin}, {Jobin}, {King}, {Sanchez-Lavega}, {Toni}, {Rimola} and
  {Poppe}}]{Drolshagen2021AA}
\bibinfo{author}{{Drolshagen}, E.}, \bibinfo{author}{{Ott}, T.},
  \bibinfo{author}{{Koschny}, D.}, \bibinfo{author}{{Drolshagen}, G.},
  \bibinfo{author}{{Vaubaillon}, J.}, \bibinfo{author}{{Colas}, F.},
  \bibinfo{author}{{Zanda}, B.}, \bibinfo{author}{{Bouley}, S.},
  \bibinfo{author}{{Jeanne}, S.}, \bibinfo{author}{{Malgoyre}, A.},
  \bibinfo{author}{{Birlan}, M.}, \bibinfo{author}{{Vernazza}, P.},
  \bibinfo{author}{{Gardiol}, D.}, \bibinfo{author}{{Nedelcu}, D.A.},
  \bibinfo{author}{{Rowe}, J.}, \bibinfo{author}{{Forcier}, M.},
  \bibinfo{author}{{Trigo-Rodriguez}, J.M.},
  \bibinfo{author}{{Pe{\~n}a-Asensio}, E.}, \bibinfo{author}{{Lamy}, H.},
  \bibinfo{author}{{Ferri{\`e}re}, L.}, \bibinfo{author}{{Barghini}, D.},
  \bibinfo{author}{{Carbognani}, A.}, \bibinfo{author}{{Di Martino}, M.},
  \bibinfo{author}{{Rasetti}, S.}, \bibinfo{author}{{Valsecchi}, G.B.},
  \bibinfo{author}{{Volpicelli}, C.A.}, \bibinfo{author}{{Di Carlo}, M.},
  \bibinfo{author}{{Knapic}, C.}, \bibinfo{author}{{Pratesi}, G.},
  \bibinfo{author}{{Riva}, W.}, \bibinfo{author}{{Stirpe}, G.M.},
  \bibinfo{author}{{Zorba}, S.}, \bibinfo{author}{{Hernandez}, O.},
  \bibinfo{author}{{Grandchamps}, A.}, \bibinfo{author}{{Jehin}, E.},
  \bibinfo{author}{{Jobin}, M.}, \bibinfo{author}{{King}, A.},
  \bibinfo{author}{{Sanchez-Lavega}, A.}, \bibinfo{author}{{Toni}, A.},
  \bibinfo{author}{{Rimola}, A.}, \bibinfo{author}{{Poppe}, B.},
  \bibinfo{year}{2021}.
\newblock \bibinfo{title}{{Luminous efficiency based on FRIPON meteors and
  limitations of ablation models}}.
\newblock \bibinfo{journal}{Astronomy and Astrophysics} \bibinfo{volume}{650},
  \bibinfo{pages}{A159}.
\newblock \DOIprefix\doi{10.1051/0004-6361/202040204}.
\bibitem[{{Dumitru} et~al.(2017){Dumitru}, {Birlan}, {Popescu} and
  {Nedelcu}}]{Dumitru2017AA}
\bibinfo{author}{{Dumitru}, B.A.}, \bibinfo{author}{{Birlan}, M.},
  \bibinfo{author}{{Popescu}, M.}, \bibinfo{author}{{Nedelcu}, D.A.},
  \bibinfo{year}{2017}.
\newblock \bibinfo{title}{{Association between meteor showers and asteroids
  using multivariate criteria}}.
\newblock \bibinfo{journal}{Astronomy and Astrophysics} \bibinfo{volume}{607},
  \bibinfo{pages}{A5}.
\newblock \DOIprefix\doi{10.1051/0004-6361/201730813}.
\bibitem[{Galindo and Lorena(2018)}]{galindo18}
\bibinfo{author}{Galindo, Y.}, \bibinfo{author}{Lorena, A.C.},
  \bibinfo{year}{2018}.
\newblock \bibinfo{title}{Deep {Transfer} {Learning} for {Meteor} {Detection}},
  in: \bibinfo{booktitle}{Anais do {XV} {Encontro} {Nacional} de
  {Inteligência} {Artificial} e {Computacional} ({ENIAC} 2018)},
  \bibinfo{publisher}{Sociedade Brasileira de Computação - SBC},
  \bibinfo{address}{São Paulo}. pp. \bibinfo{pages}{528--537}.
\newblock \URLprefix
  \url{http://portaldeconteudo.sbc.org.br/index.php/eniac/article/view/4445},
  \DOIprefix\doi{10.5753/eniac.2018.4445}.
\bibitem[{Ganju et~al.(2023)Ganju, Hatua, Jenniskens, Krishna, Ren and
  Ambardar}]{Ganju2023}
\bibinfo{author}{Ganju, S.}, \bibinfo{author}{Hatua, A.},
  \bibinfo{author}{Jenniskens, P.}, \bibinfo{author}{Krishna, S.},
  \bibinfo{author}{Ren, C.}, \bibinfo{author}{Ambardar, S.},
  \bibinfo{year}{2023}.
\newblock \bibinfo{title}{Ai-enhanced data processing and discovery crowd
  sourcing for meteor shower mapping}.
\newblock \URLprefix \url{https://arxiv.org/abs/2308.02664},
  \DOIprefix\doi{10.48550/ARXIV.2308.02664}.
\bibitem[{{Gu} et~al.(2018){Gu}, {Wang}, {Kuen}, {Ma}, {Shahroudy}, {Shuai},
  {Liu}, {Wang}, {Wang}, {Cai} and {Chen}}]{Gu2018PatRe}
\bibinfo{author}{{Gu}, J.}, \bibinfo{author}{{Wang}, Z.},
  \bibinfo{author}{{Kuen}, J.}, \bibinfo{author}{{Ma}, L.},
  \bibinfo{author}{{Shahroudy}, A.}, \bibinfo{author}{{Shuai}, B.},
  \bibinfo{author}{{Liu}, T.}, \bibinfo{author}{{Wang}, X.},
  \bibinfo{author}{{Wang}, G.}, \bibinfo{author}{{Cai}, J.},
  \bibinfo{author}{{Chen}, T.}, \bibinfo{year}{2018}.
\newblock \bibinfo{title}{{Recent advances in convolutional neural networks}}.
\newblock \bibinfo{journal}{Pattern Recognition} \bibinfo{volume}{77},
  \bibinfo{pages}{354--377}.
\newblock \DOIprefix\doi{10.1016/j.patcog.2017.10.013}.
\bibitem[{{Gural} and {{\v{S}}egon}(2009)}]{Gural2009JIMO}
\bibinfo{author}{{Gural}, P.}, \bibinfo{author}{{{\v{S}}egon}, D.},
  \bibinfo{year}{2009}.
\newblock \bibinfo{title}{{A new meteor detection processing approach for
  observations collected by the Croatian Meteor Network (CMN)}}.
\newblock \bibinfo{journal}{WGN, Journal of the International Meteor
  Organization} \bibinfo{volume}{37}, \bibinfo{pages}{28--32}.
\bibitem[{{Gural}(2011)}]{Gural2011pimo}
\bibinfo{author}{{Gural}, P.S.}, \bibinfo{year}{2011}.
\newblock \bibinfo{title}{{The California All-sky Meteor Surveillance (CAMS)
  System}}, in: \bibinfo{booktitle}{Proceedings of the International Meteor
  Conference, 29th IMC, Armagh, Northern Ireland, 2010}, pp.
  \bibinfo{pages}{28--31}.
\bibitem[{{Gural}(2019)}]{Gural2019}
\bibinfo{author}{{Gural}, P.S.}, \bibinfo{year}{2019}.
\newblock \bibinfo{title}{{Deep learning algorithms applied to the
  classification of video meteor detections}}.
\newblock \bibinfo{journal}{Monthly Notices of the Royal Astronomical Society}
  \bibinfo{volume}{489}, \bibinfo{pages}{5109--5118}.
\newblock \DOIprefix\doi{10.1093/mnras/stz2456}.
\bibitem[{{Hajdukova} et~al.(2020){Hajdukova}, {Sterken}, {Wiegert} and
  {Korno{\v{s}}}}]{Hajdukova2020PSS}
\bibinfo{author}{{Hajdukova}, M.}, \bibinfo{author}{{Sterken}, V.},
  \bibinfo{author}{{Wiegert}, P.}, \bibinfo{author}{{Korno{\v{s}}}, L.},
  \bibinfo{year}{2020}.
\newblock \bibinfo{title}{{The challenge of identifying interstellar meteors}}.
\newblock \bibinfo{journal}{Planetary and Space Science} \bibinfo{volume}{192},
  \bibinfo{pages}{105060}.
\newblock \DOIprefix\doi{10.1016/j.pss.2020.105060}.
\bibitem[{Hastie et~al.(2001)Hastie, Friedman and Tibshirani}]{Hastie2001}
\bibinfo{author}{Hastie, T.}, \bibinfo{author}{Friedman, J.},
  \bibinfo{author}{Tibshirani, R.}, \bibinfo{year}{2001}.
\newblock \bibinfo{title}{The Elements of Statistical Learning}.
\newblock \bibinfo{publisher}{Springer New York}.
\newblock \URLprefix \url{https://doi.org/10.1007/978-0-387-21606-5},
  \DOIprefix\doi{10.1007/978-0-387-21606-5}.
\bibitem[{{He} et~al.(2015){He}, {Zhang}, {Ren} and {Sun}}]{He2015arXiv}
\bibinfo{author}{{He}, K.}, \bibinfo{author}{{Zhang}, X.},
  \bibinfo{author}{{Ren}, S.}, \bibinfo{author}{{Sun}, J.},
  \bibinfo{year}{2015}.
\newblock \bibinfo{title}{{Deep Residual Learning for Image Recognition}}.
\newblock \bibinfo{journal}{arXiv e-prints} ,
  \bibinfo{pages}{arXiv:1512.03385}\DOIprefix\doi{10.48550/arXiv.1512.03385},
  \href{http://arxiv.org/abs/1512.03385}{{\tt arXiv:1512.03385}}.
\bibitem[{{Howie} et~al.(2017){Howie}, {Paxman}, {Bland}, {Towner}, {Cupak},
  {Sansom} and {Devillepoix}}]{Howie2017ExA}
\bibinfo{author}{{Howie}, R.M.}, \bibinfo{author}{{Paxman}, J.},
  \bibinfo{author}{{Bland}, P.A.}, \bibinfo{author}{{Towner}, M.C.},
  \bibinfo{author}{{Cupak}, M.}, \bibinfo{author}{{Sansom}, E.K.},
  \bibinfo{author}{{Devillepoix}, H.A.R.}, \bibinfo{year}{2017}.
\newblock \bibinfo{title}{{How to build a continental scale fireball camera
  network}}.
\newblock \bibinfo{journal}{Experimental Astronomy} \bibinfo{volume}{43},
  \bibinfo{pages}{237--266}.
\newblock \DOIprefix\doi{10.1007/s10686-017-9532-7}.
\bibitem[{{Hughes}(1978)}]{Hughes1978book}
\bibinfo{author}{{Hughes}, D.W.}, \bibinfo{year}{1978}.
\newblock \bibinfo{title}{{Meteors.}}, in: \bibinfo{editor}{{McDonnell},
  J.A.M.} (Ed.), \bibinfo{booktitle}{Cosmic Dust}, pp.
  \bibinfo{pages}{123--185}.
\bibitem[{{Jenniskens}(2017)}]{Jenniskens2017PSS}
\bibinfo{author}{{Jenniskens}, P.}, \bibinfo{year}{2017}.
\newblock \bibinfo{title}{{Meteor showers in review}}.
\newblock \bibinfo{journal}{Planetary and Space Science} \bibinfo{volume}{143},
  \bibinfo{pages}{116--124}.
\newblock \DOIprefix\doi{10.1016/j.pss.2017.01.008}.
\bibitem[{Jenniskens et~al.(2011)Jenniskens, Gural, Dynneson, Grigsby, Newman,
  Borden, Koop and Holman}]{JENNISKENS2011cams}
\bibinfo{author}{Jenniskens, P.}, \bibinfo{author}{Gural, P.},
  \bibinfo{author}{Dynneson, L.}, \bibinfo{author}{Grigsby, B.},
  \bibinfo{author}{Newman, K.}, \bibinfo{author}{Borden, M.},
  \bibinfo{author}{Koop, M.}, \bibinfo{author}{Holman, D.},
  \bibinfo{year}{2011}.
\newblock \bibinfo{title}{Cams: Cameras for allsky meteor surveillance to
  establish minor meteor showers}.
\newblock \bibinfo{journal}{Icarus} \bibinfo{volume}{216},
  \bibinfo{pages}{40--61}.
\newblock \URLprefix
  \url{https://www.sciencedirect.com/science/article/pii/S0019103511003290},
  \DOIprefix\doi{https://doi.org/10.1016/j.icarus.2011.08.012}.
\bibitem[{{Jenniskens} et~al.(2011){Jenniskens}, {Gural}, {Dynneson},
  {Grigsby}, {Newman}, {Borden}, {Koop} and {Holman}}]{Jenniskens2011Icar}
\bibinfo{author}{{Jenniskens}, P.}, \bibinfo{author}{{Gural}, P.S.},
  \bibinfo{author}{{Dynneson}, L.}, \bibinfo{author}{{Grigsby}, B.J.},
  \bibinfo{author}{{Newman}, K.E.}, \bibinfo{author}{{Borden}, M.},
  \bibinfo{author}{{Koop}, M.}, \bibinfo{author}{{Holman}, D.},
  \bibinfo{year}{2011}.
\newblock \bibinfo{title}{{CAMS: Cameras for Allsky Meteor Surveillance to
  establish minor meteor showers}}.
\newblock \bibinfo{journal}{Icarus} \bibinfo{volume}{216},
  \bibinfo{pages}{40--61}.
\newblock \DOIprefix\doi{10.1016/j.icarus.2011.08.012}.
\bibitem[{{Jiang} et~al.(2021){Jiang}, {Zhang}, {Hou}, {Cheng} and
  {Wei}}]{Jiang2021ITIP}
\bibinfo{author}{{Jiang}, P.T.}, \bibinfo{author}{{Zhang}, C.B.},
  \bibinfo{author}{{Hou}, Q.}, \bibinfo{author}{{Cheng}, M.M.},
  \bibinfo{author}{{Wei}, Y.}, \bibinfo{year}{2021}.
\newblock \bibinfo{title}{{LayerCAM: Exploring Hierarchical Class Activation
  Maps for Localization}}.
\newblock \bibinfo{journal}{IEEE Transactions on Image Processing}
  \bibinfo{volume}{30}, \bibinfo{pages}{5875--5888}.
\newblock \DOIprefix\doi{10.1109/TIP.2021.3089943}.
\bibitem[{{Jopek} and {Williams}(2013)}]{Jopek2013MNRAS}
\bibinfo{author}{{Jopek}, T.J.}, \bibinfo{author}{{Williams}, I.P.},
  \bibinfo{year}{2013}.
\newblock \bibinfo{title}{{Stream and sporadic meteoroids associated with
  near-Earth objects}}.
\newblock \bibinfo{journal}{Monthly Notices of the Royal Astronomical Society}
  \bibinfo{volume}{430}, \bibinfo{pages}{2377--2389}.
\newblock \DOIprefix\doi{10.1093/mnras/stt057}.
\bibitem[{{Korno{\v{s}}} et~al.(2014){Korno{\v{s}}}, {Koukal}, {Piffl} and
  {T{\'o}th}}]{Kornos2014pim3}
\bibinfo{author}{{Korno{\v{s}}}, L.}, \bibinfo{author}{{Koukal}, J.},
  \bibinfo{author}{{Piffl}, R.}, \bibinfo{author}{{T{\'o}th}, J.},
  \bibinfo{year}{2014}.
\newblock \bibinfo{title}{{EDMOND Meteor Database}}, in:
  \bibinfo{editor}{{Gyssens}, M.}, \bibinfo{editor}{{Roggemans}, P.},
  \bibinfo{editor}{{Zoladek}, P.} (Eds.), \bibinfo{booktitle}{Proceedings of
  the International Meteor Conference, Poznan, Poland, 22-25 August 2013}, pp.
  \bibinfo{pages}{23--25}.
\bibitem[{{Koschny} et~al.(2017){Koschny}, {Drolshagen}, {Drolshagen},
  {Kretschmer}, {Ott}, {Drolshagen} and {Poppe}}]{Koschny2017PSS}
\bibinfo{author}{{Koschny}, D.}, \bibinfo{author}{{Drolshagen}, E.},
  \bibinfo{author}{{Drolshagen}, S.}, \bibinfo{author}{{Kretschmer}, J.},
  \bibinfo{author}{{Ott}, T.}, \bibinfo{author}{{Drolshagen}, G.},
  \bibinfo{author}{{Poppe}, B.}, \bibinfo{year}{2017}.
\newblock \bibinfo{title}{{Flux densities of meteoroids derived from optical
  double-station observations}}.
\newblock \bibinfo{journal}{Planetary and Space Science} \bibinfo{volume}{143},
  \bibinfo{pages}{230--237}.
\newblock \DOIprefix\doi{10.1016/j.pss.2016.12.007}.
\bibitem[{{Koschny} et~al.(2019){Koschny}, {Soja}, {Engrand}, {Flynn}, {Lasue},
  {Levasseur-Regourd}, {Malaspina}, {Nakamura}, {Poppe}, {Sterken} and
  {Trigo-Rodr{\'\i}guez}}]{Koschny2019SSRv}
\bibinfo{author}{{Koschny}, D.}, \bibinfo{author}{{Soja}, R.H.},
  \bibinfo{author}{{Engrand}, C.}, \bibinfo{author}{{Flynn}, G.J.},
  \bibinfo{author}{{Lasue}, J.}, \bibinfo{author}{{Levasseur-Regourd}, A.C.},
  \bibinfo{author}{{Malaspina}, D.}, \bibinfo{author}{{Nakamura}, T.},
  \bibinfo{author}{{Poppe}, A.R.}, \bibinfo{author}{{Sterken}, V.J.},
  \bibinfo{author}{{Trigo-Rodr{\'\i}guez}, J.M.}, \bibinfo{year}{2019}.
\newblock \bibinfo{title}{{Interplanetary Dust, Meteoroids, Meteors and
  Meteorites}}.
\newblock \bibinfo{journal}{Space Science Reviews} \bibinfo{volume}{215},
  \bibinfo{pages}{34}.
\newblock \DOIprefix\doi{10.1007/s11214-019-0597-7}.
\bibitem[{{Koten} et~al.(2019){Koten}, {Rendtel}, {Shrben{\'y}}, {Gural},
  {Borovi{\v{c}}ka} and {Kozak}}]{Koten2019msme}
\bibinfo{author}{{Koten}, P.}, \bibinfo{author}{{Rendtel}, J.},
  \bibinfo{author}{{Shrben{\'y}}, L.}, \bibinfo{author}{{Gural}, P.},
  \bibinfo{author}{{Borovi{\v{c}}ka}, J.}, \bibinfo{author}{{Kozak}, P.},
  \bibinfo{year}{2019}.
\newblock \bibinfo{title}{{Meteors and Meteor Showers as Observed by Optical
  Techniques}}, in: \bibinfo{editor}{{Ryabova}, G.O.},
  \bibinfo{editor}{{Asher}, D.J.}, \bibinfo{editor}{{Campbell-Brown}, M.J.}
  (Eds.), \bibinfo{booktitle}{Meteoroids: Sources of Meteors on Earth and
  Beyond}, p.~\bibinfo{pages}{90}.
\newblock \DOIprefix\doi{Here DOI}.
\bibitem[{Lauretta and McSween(2006)}]{Walker2006book}
\bibinfo{editor}{Lauretta, D.S.}, \bibinfo{editor}{McSween, H.Y.} (Eds.),
  \bibinfo{year}{2006}.
\newblock \bibinfo{title}{Meteorites and the Early Solar System {II}}.
\newblock \bibinfo{publisher}{University of Arizona Press}.
\newblock \URLprefix \url{https://doi.org/10.2307/j .ctv1v7zdmm},
  \DOIprefix\doi{10.2307/j.ctv1v7zdmm}.
\bibitem[{Le~Lan and Dinh(2021)}]{LeLan21}
\bibinfo{author}{Le~Lan, C.}, \bibinfo{author}{Dinh, L.}, \bibinfo{year}{2021}.
\newblock \bibinfo{title}{Perfect {Density} {Models} {Cannot} {Guarantee}
  {Anomaly} {Detection}}.
\newblock \bibinfo{journal}{Entropy} \bibinfo{volume}{23},
  \bibinfo{pages}{1690}.
\newblock \URLprefix \url{https://www.mdpi.com/1099-4300/23/12/1690},
  \DOIprefix\doi{10.3390/e23121690}.
\bibitem[{Marsola and Lorena(2019)}]{Marsola19}
\bibinfo{author}{Marsola, T.C.L.}, \bibinfo{author}{Lorena, A.C.},
  \bibinfo{year}{2019}.
\newblock \bibinfo{title}{Meteor detection using deep convolutional neural
  networks}.
\newblock \bibinfo{journal}{Anais do 14º Simp{\'o}sio Brasileiro de
  Automaç{\~a}o Inteligente} .
\bibitem[{{Molau}(2001)}]{Molau2001ESASP}
\bibinfo{author}{{Molau}, S.}, \bibinfo{year}{2001}.
\newblock \bibinfo{title}{{The AKM video meteor network}}, in:
  \bibinfo{editor}{{Warmbein}, B.} (Ed.), \bibinfo{booktitle}{Meteoroids 2001
  Conference}, pp. \bibinfo{pages}{315--318}.
\bibitem[{{Molau} and {Gural}(2005)}]{Molau2005JIMO}
\bibinfo{author}{{Molau}, S.}, \bibinfo{author}{{Gural}, P.S.},
  \bibinfo{year}{2005}.
\newblock \bibinfo{title}{{A review of video meteor detection and analysis
  software}}.
\newblock \bibinfo{journal}{WGN, Journal of the International Meteor
  Organization} \bibinfo{volume}{33}, \bibinfo{pages}{15--20}.
\bibitem[{{Nedelcu} et~al.(2018){Nedelcu}, {Birlan}, {Turcu}, {Badescu},
  {Boaca}, {Gornea}, {Blagoi}, {Danescu} and {Paraschiv}}]{Nedelcu18moroi}
\bibinfo{author}{{Nedelcu}, A.D.}, \bibinfo{author}{{Birlan}, M.},
  \bibinfo{author}{{Turcu}, V.}, \bibinfo{author}{{Badescu}, O.},
  \bibinfo{author}{{Boaca}, I.}, \bibinfo{author}{{Gornea}, A.},
  \bibinfo{author}{{Blagoi}, O.}, \bibinfo{author}{{Danescu}, C.},
  \bibinfo{author}{{Paraschiv}, P.}, \bibinfo{year}{2018}.
\newblock \bibinfo{title}{{The MOROI network. Meteorites Orbits Reconstruction
  by Optical Imaging}}, in: \bibinfo{booktitle}{ZAC 2018 - International
  Conference Outlook in Astronomy}, p.~\bibinfo{pages}{21}.
\bibitem[{{Nikolic}(2019)}]{Nikolic2019}
\bibinfo{author}{{Nikolic}, V.}, \bibinfo{year}{2019}.
\newblock \bibinfo{title}{{Automation of a video meteor network}}, in:
  \bibinfo{editor}{{Rudawska}, R.}, \bibinfo{editor}{{Rendtel}, J.},
  \bibinfo{editor}{{Powell}, C.}, \bibinfo{editor}{{Lunsford}, R.},
  \bibinfo{editor}{{Verbeeck}, C.}, \bibinfo{editor}{{Knofel}, A.} (Eds.),
  \bibinfo{booktitle}{International Meteor Conference, Pezinok-Modra,
  Slovakia}, pp. \bibinfo{pages}{139--140}.
\bibitem[{{Pe{\~n}a-Asensio} et~al.(2021a){Pe{\~n}a-Asensio},
  {Trigo-Rodr{\'\i}guez}, {Gritsevich} and {Rimola}}]{Eloy2021MNRAS}
\bibinfo{author}{{Pe{\~n}a-Asensio}, E.},
  \bibinfo{author}{{Trigo-Rodr{\'\i}guez}, J.M.},
  \bibinfo{author}{{Gritsevich}, M.}, \bibinfo{author}{{Rimola}, A.},
  \bibinfo{year}{2021}a.
\newblock \bibinfo{title}{{Accurate 3D fireball trajectory and orbit
  calculation using the 3D-FIRETOC automatic Python code}}.
\newblock \bibinfo{journal}{Monthly Notices of the Royal Astronomical Society}
  \bibinfo{volume}{504}, \bibinfo{pages}{4829--4840}.
\newblock \DOIprefix\doi{10.1093/mnras/stab999},
  \href{http://arxiv.org/abs/2103.13758}{{\tt arXiv:2103.13758}}.
\bibitem[{{Pe{\~n}a-Asensio} et~al.(2021b){Pe{\~n}a-Asensio},
  {Trigo-Rodr{\'\i}guez}, {Langbroek}, {Rimola} and
  {Robles}}]{Eloy2021Astrodyn}
\bibinfo{author}{{Pe{\~n}a-Asensio}, E.},
  \bibinfo{author}{{Trigo-Rodr{\'\i}guez}, J.M.}, \bibinfo{author}{{Langbroek},
  M.}, \bibinfo{author}{{Rimola}, A.}, \bibinfo{author}{{Robles}, A.J.},
  \bibinfo{year}{2021}b.
\newblock \bibinfo{title}{{Using fireball networks to track more frequent
  reentries: Falcon 9 upper-stage orbit determination from video recordings}}.
\newblock \bibinfo{journal}{Astrodynamics} \bibinfo{volume}{5},
  \bibinfo{pages}{347--358}.
\newblock \DOIprefix\doi{10.1007/s42064-021-0112-2},
  \href{http://arxiv.org/abs/2109.01004}{{\tt arXiv:2109.01004}}.
\bibitem[{{Pe{\~n}a-Asensio} et~al.(2022){Pe{\~n}a-Asensio},
  {Trigo-Rodr{\'\i}guez} and {Rimola}}]{Eloy2022AJ}
\bibinfo{author}{{Pe{\~n}a-Asensio}, E.},
  \bibinfo{author}{{Trigo-Rodr{\'\i}guez}, J.M.}, \bibinfo{author}{{Rimola},
  A.}, \bibinfo{year}{2022}.
\newblock \bibinfo{title}{{Orbital Characterization of Superbolides Observed
  from Space: Dynamical Association with Near-Earth Objects, Meteoroid Streams,
  and Identification of Hyperbolic Meteoroids}}.
\newblock \bibinfo{journal}{The Astronomical Journal} \bibinfo{volume}{164},
  \bibinfo{pages}{76}.
\newblock \DOIprefix\doi{10.3847/1538-3881/ac75d2},
  \href{http://arxiv.org/abs/2206.03115}{{\tt arXiv:2206.03115}}.
\bibitem[{{Pe{\~n}a-Asensio} et~al.(2023){Pe{\~n}a-Asensio},
  {Trigo-Rodr{\'\i}guez}, {Rimola}, {Corretg{\'e}-Gilart} and
  {Koschny}}]{Eloy2023MNRAS}
\bibinfo{author}{{Pe{\~n}a-Asensio}, E.},
  \bibinfo{author}{{Trigo-Rodr{\'\i}guez}, J.M.}, \bibinfo{author}{{Rimola},
  A.}, \bibinfo{author}{{Corretg{\'e}-Gilart}, M.}, \bibinfo{author}{{Koschny},
  D.}, \bibinfo{year}{2023}.
\newblock \bibinfo{title}{{Identifying meteorite droppers among the population
  of bright 'sporadic' bolides imaged by the Spanish Meteor Network during the
  spring of 2022}}.
\newblock \bibinfo{journal}{Monthly Notices of the Royal Astronomical Society}
  \bibinfo{volume}{520}, \bibinfo{pages}{5173--5182}.
\newblock \DOIprefix\doi{10.1093/mnras/stad102},
  \href{http://arxiv.org/abs/2301.03515}{{\tt arXiv:2301.03515}}.
\bibitem[{Rohwer and Laurie(2006)}]{RohwerLaurie06}
\bibinfo{author}{Rohwer, C.H.}, \bibinfo{author}{Laurie, D.P.},
  \bibinfo{year}{2006}.
\newblock \bibinfo{title}{The discrete pulse transform}.
\newblock \bibinfo{journal}{SIAM J. Math. Anal.} \bibinfo{volume}{38},
  \bibinfo{pages}{1012--1034}.
\bibitem[{Selvaraju et~al.(2016)Selvaraju, Das, Vedantam, Cogswell, Parikh and
  Batra}]{Selvaraju2016}
\bibinfo{author}{Selvaraju, R.R.}, \bibinfo{author}{Das, A.},
  \bibinfo{author}{Vedantam, R.}, \bibinfo{author}{Cogswell, M.},
  \bibinfo{author}{Parikh, D.}, \bibinfo{author}{Batra, D.},
  \bibinfo{year}{2016}.
\newblock \bibinfo{title}{Grad-cam: Why did you say that?}
\newblock \URLprefix \url{https://arxiv.org/abs/1611.07450},
  \DOIprefix\doi{10.48550/ARXIV.1611.07450}.
\bibitem[{{Sennlaub} et~al.(2022){Sennlaub}, {Hofmann}, {Hankey}, {Ennes},
  {M{\"u}ller}, {Kroll} and {M{\"a}der}}]{Sennlaub22}
\bibinfo{author}{{Sennlaub}, R.}, \bibinfo{author}{{Hofmann}, M.},
  \bibinfo{author}{{Hankey}, M.}, \bibinfo{author}{{Ennes}, M.},
  \bibinfo{author}{{M{\"u}ller}, T.}, \bibinfo{author}{{Kroll}, P.},
  \bibinfo{author}{{M{\"a}der}, P.}, \bibinfo{year}{2022}.
\newblock \bibinfo{title}{{Object classification on video data of meteors and
  meteor-like phenomena: algorithm and data}}.
\newblock \bibinfo{journal}{Monthly Notices of the Royal Astronomical Society}
  \bibinfo{volume}{516}, \bibinfo{pages}{811--823}.
\newblock \DOIprefix\doi{10.1093/mnras/stac1948},
  \href{http://arxiv.org/abs/2208.14914}{{\tt arXiv:2208.14914}}.
\bibitem[{{Siladi} et~al.(2015){Siladi}, {Vida} and
  {Nyarko}}]{Siladji2015moconf}
\bibinfo{author}{{Siladi}, E.}, \bibinfo{author}{{Vida}, D.},
  \bibinfo{author}{{Nyarko}, K.}, \bibinfo{year}{2015}.
\newblock \bibinfo{title}{{Video meteor detection filtering using soft
  computing methods}}, in: \bibinfo{booktitle}{International Meteor Conference
  Mistelbach, Austria}, p.~\bibinfo{pages}{24}.
\bibitem[{{Silber} et~al.(2018){Silber}, {Boslough}, {Hocking}, {Gritsevich}
  and {Whitaker}}]{Silber2018AdSpR}
\bibinfo{author}{{Silber}, E.A.}, \bibinfo{author}{{Boslough}, M.},
  \bibinfo{author}{{Hocking}, W.K.}, \bibinfo{author}{{Gritsevich}, M.},
  \bibinfo{author}{{Whitaker}, R.W.}, \bibinfo{year}{2018}.
\newblock \bibinfo{title}{{Physics of meteor generated shock waves in the
  Earth's atmosphere - A review}}.
\newblock \bibinfo{journal}{Advances in Space Research} \bibinfo{volume}{62},
  \bibinfo{pages}{489--532}.
\newblock \DOIprefix\doi{10.1016/j.asr.2018.05.010},
  \href{http://arxiv.org/abs/1805.07842}{{\tt arXiv:1805.07842}}.
\bibitem[{Simonyan and Zisserman(2015)}]{simonyan15vgg}
\bibinfo{author}{Simonyan, K.}, \bibinfo{author}{Zisserman, A.},
  \bibinfo{year}{2015}.
\newblock \bibinfo{title}{Very {Deep} {Convolutional} {Networks} for
  {Large}-{Scale} {Image} {Recognition}}.
\newblock \URLprefix \url{http://arxiv.org/abs/1409.1556}.
  \bibinfo{note}{arXiv:1409.1556 [cs]}.
\bibitem[{{SonotaCo}(2016)}]{SonotaCo2016JIMO}
\bibinfo{author}{{SonotaCo}}, \bibinfo{year}{2016}.
\newblock \bibinfo{title}{{Observation error propagation on video meteor orbit
  determination}}.
\newblock \bibinfo{journal}{WGN, Journal of the International Meteor
  Organization} \bibinfo{volume}{44}, \bibinfo{pages}{42--45}.
\bibitem[{{Spurn{\'y}} et~al.(2007){Spurn{\'y}}, {Borovi{\v{c}}ka} and
  {Shrben{\'y}}}]{Spurny2007IAUS}
\bibinfo{author}{{Spurn{\'y}}, P.}, \bibinfo{author}{{Borovi{\v{c}}ka}, J.},
  \bibinfo{author}{{Shrben{\'y}}, L.}, \bibinfo{year}{2007}.
\newblock \bibinfo{title}{{Automation of the Czech part of the European
  fireball network: equipment, methods and first results}}, in:
  \bibinfo{editor}{{Valsecchi}, G.B.}, \bibinfo{editor}{{Vokrouhlick{\'y}},
  D.}, \bibinfo{editor}{{Milani}, A.} (Eds.), \bibinfo{booktitle}{Near Earth
  Objects, our Celestial Neighbors: Opportunity and Risk}, pp.
  \bibinfo{pages}{121--130}.
\newblock \DOIprefix\doi{10.1017/S1743921307003146}.
\bibitem[{{Subasinghe} et~al.(2017){Subasinghe}, {Campbell-Brown} and
  {Stokan}}]{Subasinghe2017PSS}
\bibinfo{author}{{Subasinghe}, D.}, \bibinfo{author}{{Campbell-Brown}, M.},
  \bibinfo{author}{{Stokan}, E.}, \bibinfo{year}{2017}.
\newblock \bibinfo{title}{{Luminous efficiency estimates of meteors -I.
  Uncertainty analysis}}.
\newblock \bibinfo{journal}{Planetary and Space Science} \bibinfo{volume}{143},
  \bibinfo{pages}{71--77}.
\newblock \DOIprefix\doi{10.1016/j.pss.2016.12.009},
  \href{http://arxiv.org/abs/1704.08656}{{\tt arXiv:1704.08656}}.
\bibitem[{{Suk} and {{\v{S}}imberov{\'a}}(2017)}]{Suk2017EMP}
\bibinfo{author}{{Suk}, T.}, \bibinfo{author}{{{\v{S}}imberov{\'a}}, S.},
  \bibinfo{year}{2017}.
\newblock \bibinfo{title}{{Automated Meteor Detection by All-Sky Digital Camera
  Systems}}.
\newblock \bibinfo{journal}{Earth Moon and Planets} \bibinfo{volume}{120},
  \bibinfo{pages}{189--215}.
\newblock \DOIprefix\doi{10.1007/s11038-017-9511-z}.
\bibitem[{{Trigo-Rodr{\'\i}guez}(2019)}]{Trigo2019hmep}
\bibinfo{author}{{Trigo-Rodr{\'\i}guez}, J.M.}, \bibinfo{year}{2019}.
\newblock \bibinfo{title}{{The flux of meteoroids over time: meteor emission
  spectroscopy and the delivery of volatiles and chondritic materials to
  Earth}}, in: \bibinfo{booktitle}{Hypersonic Meteoroid Entry Physics}.
  \bibinfo{publisher}{{IOP} Publishing}, p.~\bibinfo{pages}{4}.
\newblock \DOIprefix\doi{10.1088/2053-2563/aae894ch4}.
\bibitem[{Trigo-Rodr{\'\i}guez(2022)}]{Trigo2022}
\bibinfo{author}{Trigo-Rodr{\'\i}guez, J.M.}, \bibinfo{year}{2022}.
\newblock \bibinfo{title}{Asteroid Impact Risk: Impact Hazard from Asteroids
  and Comets}.
\newblock \bibinfo{publisher}{Springer Nature}.
\bibitem[{{Trigo-Rodr{\'\i}guez} and {Blum}(2022)}]{Trigo2022MNRAS}
\bibinfo{author}{{Trigo-Rodr{\'\i}guez}, J.M.}, \bibinfo{author}{{Blum}, J.},
  \bibinfo{year}{2022}.
\newblock \bibinfo{title}{{Learning about comets from the study of mass
  distributions and fluxes of meteoroid streams}}.
\newblock \bibinfo{journal}{Monthly Notices of the Royal Astronomical Society}
  \bibinfo{volume}{512}, \bibinfo{pages}{2277--2289}.
\newblock \DOIprefix\doi{10.1093/mnras/stab2827},
  \href{http://arxiv.org/abs/2109.14428}{{\tt arXiv:2109.14428}}.
\bibitem[{{Trigo-Rodr{\'\i}guez} et~al.(2006){Trigo-Rodr{\'\i}guez}, {Llorca},
  {Castro-Tirado}, {Ortiz}, {Docobo} and {Fabregat}}]{Trigo2006AG}
\bibinfo{author}{{Trigo-Rodr{\'\i}guez}, J.M.}, \bibinfo{author}{{Llorca}, J.},
  \bibinfo{author}{{Castro-Tirado}, A.J.}, \bibinfo{author}{{Ortiz}, J.L.},
  \bibinfo{author}{{Docobo}, J.A.}, \bibinfo{author}{{Fabregat}, J.},
  \bibinfo{year}{2006}.
\newblock \bibinfo{title}{{The Spanish fireball network}}.
\newblock \bibinfo{journal}{Astronomy and Geophysics} \bibinfo{volume}{47},
  \bibinfo{pages}{6.26--6.28}.
\newblock \DOIprefix\doi{10.1111/j.1468-4004.2006.47626.x}.
\bibitem[{{Vaubaillon} et~al.(2019){Vaubaillon}, {Neslu{\v{s}}an}, {Sekhar},
  {Rudawska} and {Ryabova}}]{Vaubaillon2019}
\bibinfo{author}{{Vaubaillon}, J.}, \bibinfo{author}{{Neslu{\v{s}}an}, L.},
  \bibinfo{author}{{Sekhar}, A.}, \bibinfo{author}{{Rudawska}, R.},
  \bibinfo{author}{{Ryabova}, G.O.}, \bibinfo{year}{2019}.
\newblock \bibinfo{title}{{From Parent Body to Meteor Shower: The Dynamics of
  Meteoroid Streams}}, in: \bibinfo{editor}{{Ryabova}, G.O.},
  \bibinfo{editor}{{Asher}, D.J.}, \bibinfo{editor}{{Campbell-Brown}, M.J.}
  (Eds.), \bibinfo{booktitle}{Meteoroids: Sources of Meteors on Earth and
  Beyond}. \bibinfo{publisher}{Cambridge University Press.}, p.
  \bibinfo{pages}{161}.
\bibitem[{{Vida} et~al.(2021){Vida}, {{\v{S}}egon}, {Gural}, {Brown},
  {McIntyre}, {Dijkema}, {Pavleti{\'c}}, {Kuki{\'c}}, {Mazur}, {Eschman},
  {Roggemans}, {Merlak} and {Zubovi{\'c}}}]{Vida2021MNRAS}
\bibinfo{author}{{Vida}, D.}, \bibinfo{author}{{{\v{S}}egon}, D.},
  \bibinfo{author}{{Gural}, P.S.}, \bibinfo{author}{{Brown}, P.G.},
  \bibinfo{author}{{McIntyre}, M.J.M.}, \bibinfo{author}{{Dijkema}, T.J.},
  \bibinfo{author}{{Pavleti{\'c}}, L.}, \bibinfo{author}{{Kuki{\'c}}, P.},
  \bibinfo{author}{{Mazur}, M.J.}, \bibinfo{author}{{Eschman}, P.},
  \bibinfo{author}{{Roggemans}, P.}, \bibinfo{author}{{Merlak}, A.},
  \bibinfo{author}{{Zubovi{\'c}}, D.}, \bibinfo{year}{2021}.
\newblock \bibinfo{title}{{The Global Meteor Network - Methodology and first
  results}}.
\newblock \bibinfo{journal}{Monthly Notices of the Royal Astronomical Society}
  \bibinfo{volume}{506}, \bibinfo{pages}{5046--5074}.
\newblock \DOIprefix\doi{10.1093/mnras/stab2008},
  \href{http://arxiv.org/abs/2107.12335}{{\tt arXiv:2107.12335}}.
\bibitem[{V{\'{\i}}tek et~al.(2011)V{\'{\i}}tek, Fliegel, P{\'{a}}ta and
  Koten}]{Vitek11maia}
\bibinfo{author}{V{\'{\i}}tek, S.}, \bibinfo{author}{Fliegel, K.},
  \bibinfo{author}{P{\'{a}}ta, P.}, \bibinfo{author}{Koten, P.},
  \bibinfo{year}{2011}.
\newblock \bibinfo{title}{{MAIA}: Technical development of a novel system for
  video observations of meteors}.
\newblock \bibinfo{journal}{Acta Polytechnica} \bibinfo{volume}{51}.
\newblock \URLprefix \url{https://doi.org/10.14311/1340},
  \DOIprefix\doi{10.14311/1340}.
\bibitem[{{V{\'\i}tek} and {Nasyrova}(2019)}]{VitekNasyrova19}
\bibinfo{author}{{V{\'\i}tek}, S.}, \bibinfo{author}{{Nasyrova}, M.},
  \bibinfo{year}{2019}.
\newblock \bibinfo{title}{{Fast meteor tracking in noisy video sequences}}.
\newblock \bibinfo{journal}{Astronomische Nachrichten} \bibinfo{volume}{340},
  \bibinfo{pages}{646--651}.
\newblock \DOIprefix\doi{10.1002/asna.201913670}.
\bibitem[{Weryk et~al.(2013)Weryk, Campbell-Brown, Wiegert, Brown, Krzeminski
  and Musci}]{Weryk13camo}
\bibinfo{author}{Weryk, R.}, \bibinfo{author}{Campbell-Brown, M.},
  \bibinfo{author}{Wiegert, P.}, \bibinfo{author}{Brown, P.},
  \bibinfo{author}{Krzeminski, Z.}, \bibinfo{author}{Musci, R.},
  \bibinfo{year}{2013}.
\newblock \bibinfo{title}{The canadian automated meteor observatory (camo):
  System overview}.
\newblock \bibinfo{journal}{Icarus} \bibinfo{volume}{225},
  \bibinfo{pages}{614--622}.
\newblock \URLprefix
  \url{https://www.sciencedirect.com/science/article/pii/S0019103513001905},
  \DOIprefix\doi{https://doi.org/10.1016/j.icarus.2013.04.025}.
\bibitem[{{Weryk} et~al.(2013){Weryk}, {Campbell-Brown}, {Wiegert}, {Brown},
  {Krzeminski} and {Musci}}]{Weryk2013Icar}
\bibinfo{author}{{Weryk}, R.J.}, \bibinfo{author}{{Campbell-Brown}, M.D.},
  \bibinfo{author}{{Wiegert}, P.A.}, \bibinfo{author}{{Brown}, P.G.},
  \bibinfo{author}{{Krzeminski}, Z.}, \bibinfo{author}{{Musci}, R.},
  \bibinfo{year}{2013}.
\newblock \bibinfo{title}{{The Canadian Automated Meteor Observatory (CAMO):
  System overview}}.
\newblock \bibinfo{journal}{Icarus} \bibinfo{volume}{225},
  \bibinfo{pages}{614--622}.
\newblock \DOIprefix\doi{10.1016/j.icarus.2013.04.025}.
\bibitem[{{Wiegert} and {Brown}(2004)}]{Wiegert2004EMP}
\bibinfo{author}{{Wiegert}, P.}, \bibinfo{author}{{Brown}, P.},
  \bibinfo{year}{2004}.
\newblock \bibinfo{title}{{The problem of linking minor meteor showers to their
  parent bodies: initial considerations}}.
\newblock \bibinfo{journal}{Earth Moon and Planets} \bibinfo{volume}{95},
  \bibinfo{pages}{19--26}.
\newblock \DOIprefix\doi{10.1007/s11038-005-4342-8}.
\bibitem[{Xiao et~al.(2017)Xiao, Rasul and Vollgraf}]{xiao2017}
\bibinfo{author}{Xiao, H.}, \bibinfo{author}{Rasul, K.},
  \bibinfo{author}{Vollgraf, R.}, \bibinfo{year}{2017}.
\newblock \bibinfo{title}{Fashion-{MNIST}: a {Novel} {Image} {Dataset} for
  {Benchmarking} {Machine} {Learning} {Algorithms}}.
\newblock \URLprefix \url{http://arxiv.org/abs/1708.07747}.
  \bibinfo{note}{arXiv:1708.07747 [cs, stat]}.
\bibitem[{{Zhou} et~al.(2015){Zhou}, {Khosla}, {Lapedriza}, {Oliva} and
  {Torralba}}]{Zhou2015arXiv151204150Z}
\bibinfo{author}{{Zhou}, B.}, \bibinfo{author}{{Khosla}, A.},
  \bibinfo{author}{{Lapedriza}, A.}, \bibinfo{author}{{Oliva}, A.},
  \bibinfo{author}{{Torralba}, A.}, \bibinfo{year}{2015}.
\newblock \bibinfo{title}{{Learning Deep Features for Discriminative
  Localization}}.
\newblock \bibinfo{journal}{arXiv e-prints} ,
  \bibinfo{pages}{arXiv:1512.04150}\DOIprefix\doi{10.48550/arXiv.1512.04150},
  \href{http://arxiv.org/abs/1512.04150}{{\tt arXiv:1512.04150}}.

\end{thebibliography}






\end{document}